\documentclass[prx,twocolumn,superscriptaddress, amssymb, amsmath]{revtex4-2}
\usepackage{graphicx}% Include figure files
\usepackage{dcolumn}% Align table columns on decimal point
\usepackage{bm}% bold math
\usepackage{wasysym}
\usepackage{color}
\usepackage{xcolor}
\usepackage{hyperref}
\usepackage[normalem]{ulem}
\usepackage{physics}
\usepackage{relsize}

\newcommand{\beq}{\begin{equation}}
\newcommand{\eeq}{\end{equation}}
\newcommand{\beqn}{\begin{eqnarray}}
\newcommand{\eeqn}{\end{eqnarray}}

\newcommand{\ra}{\rightarrow}

\newcommand{\cA}{ {\cal A} }

\newcommand{\cC}{ {\cal C} }

\newcommand{\cE}{ {\cal E} }

\newcommand{\cH}{ {\cal H} }

\newcommand{\cL}{ {\cal L} }
\newcommand{\cP}{ {\cal P} }
\newcommand{\cS}{ {\cal S} }

\newcommand{\vect}[1]{{\bm{#1}}}
\newcommand{\Int}{\mathrm{int}}
\newcommand{\ii}{\mathrm{i}}

\newcommand{\hrho}{\hat{\rho}}
\newcommand{\htheta}{\hat{\theta}}
\newcommand{\hphi}{\hat{\phi}}
\newcommand{\hn}{\hat{n}}
\newcommand{\hO}{\hat{O}}

\newcommand{\SO}{\mathrm{SO}}

\newcommand{\U}{\mathrm{U}}

\newcommand{\tc}{\textrm{tc}}
\newcommand{\secref}[1]{Sec.\,\ref{#1}}
\newcommand{\appref}[1]{Appendix.\,\ref{#1}}
\newcommand{\eqnref}[1]{Eq.\,\eqref{#1}}
\newcommand{\figref}[1]{Fig.\,\ref{#1}}

\newcommand{\rd}{\partial}

\newcommand{\vdagger}{{\vphantom{\dagger}}}

\newcommand{\svdots}{\raisebox{3pt}{\scalebox{.75}{$\vdots$}}} % <- Works
\newcommand{\sddots}{\raisebox{3pt}{\scalebox{.75}{$\ddots$}}} % <- Do not work

\newcommand{\rAngle}{\rangle \hspace{-2pt} \rangle }
\newcommand{\lAngle}{\langle \hspace{-2pt} \langle }

\newcommand{\bsigma}{{\bm{\sigma}}}

\newcommand{\bs}{{\bm{s}}}

\newcommand{\bl}{{\bm{l}}}

\newcommand{\bmm}{{\bm{m}}}

\newcommand{\cmj}[1]{\textcolor{black}{#1}}

\renewcommand{\tr}{\mathrm{tr}}
\renewcommand{\O}{\mathrm{O}}

\begin{document}

\title{Quantum criticality under decoherence or weak measurement}

\author{Jong Yeon Lee}

\affiliation{Kavli Institute for Theoretical Physics, University of California, Santa Barbara, CA 93106}

\author{Chao-Ming Jian}

\affiliation{Department of Physics, Cornell University, Ithaca, New York 14853}

\author{Cenke Xu}

\affiliation{Department of Physics, University of California,
Santa Barbara, CA 93106}

%\date{\today}

\begin{abstract}

Decoherence inevitably happens when a quantum state is exposed to its environment, which can affect quantum critical points (QCP) in a nontrivial way. As was pointed out in recent literature on $(1+1)d$ conformal field theory (CFT)~\cite{altman}, the effect of weak measurement can be mathematically mapped to the problem of boundary CFT. In this work, we focus on the $(2+1)d$ QCPs, whose boundary and defect effects have attracted enormous theoretical and numerical interests very recently. We focus on decoherence caused by weak measurements with and without postselecting the measurement outcomes. Our main results are: (1) for an $\O(N)$ Wilson-Fisher QCP under weak measurement with postselection, an observer would in general observe two different types of boundary/defect criticality with very different behaviors from the well-known Wilson-Fisher fixed points; in particular, it is possible to observe the recently proposed exotic ``{\it extraordinary-log}" correlation. (2) An extra quantum phase transition can be driven by decoherence, if we consider quantities nonlinear with the decohered density matrix, such as the Renyi entropy. We demonstrate the connection between this transition to the information-theoretic transition driven by an error in the toric code model. (3) When there is no postselection, though correlation functions between local operators remain the same as the undecohered pure state, nonlocal operators such as the ``{\it disorder operator}" would have qualitatively distinct behaviors; and we also show that the decoherence can lead to confinement.

\end{abstract}

\maketitle

\section{Introduction}

When a quantum state is exposed to an environment, it is being constantly probed and ``measured", forming entanglement with the degrees of freedom in the environment.  If one is ignorant of the environment and the measurement outcome is lost, it amounts to tracing out the environment's degrees of freedom and the original pure quantum state becomes a mixed state, which results in the loss of coherent quantum information. This process is referred to as quantum decoherence, and it is the bridge between the quantum mechanics that governs the microscopic nature, and our classical macroscopic world~\cite{decoherence}.  More generally, if a quantum state is weakly measured and the measurement outcome is still accessible, one can consider the effect of postselecting the measurement outcome and study process that is more general than decoherence. In recent years there has been a surge of progress in simulating quantum states of matter with nontrivial entanglement on platforms summarized as the noisy intermediate-scale quantum (NISQ) technology~\cite{nisq}, including simulating exotic quantum many-body states such as topological order, spin liquids, and symmetry protected topological (SPT) states~\cite{earlyToric,googleToric,SPT2019,KZ2019,SL2021}, which have long been discussed in condensed-matter physics. In these platforms, decoherence can happen due to various reasons, which motivated recent inspection of the fate of SPT states under decoherence~\cite{nishimorilee,nishimorinat,de_Groot_2022,sptdecohere} (related studies motivated from other contexts were also conducted~\cite{mawang,qibi}). 

Quantum criticality represents another class of quantum many-body states with peculiar and universal entanglement. Recently a class of $(1+1)d$ conformal field theory (CFT), i.e., the Luttinger liquid under weak measurements has been studied, and it was pointed out that the effect of weak measurements can be mathematically mapped to the problem of the boundary of the CFT~\cite{altman}, which is a subject that was studied extensively in the past~\cite{kanefisher}. The same trick was used in the recent study of SPT states under decoherence~\cite{sptdecohere}, which exploited the observation that the wave function of the SPT states can be mapped to the partition function of the boundary states of the system after a space-time rotation~\cite{sptwf,YouXu2013}.

In this work, we focus on quantum critical points in $(2+1)d$. The connection between the decohered QCPs (or QCPs under weak measurement) in the bulk and the boundary criticality still holds, but the boundary criticality of $(2+1)d$ QCPs is a subject that has only been carefully studied very recently, and it has attracted enormous interests from both the theoretical and numerical communities~\cite{groveredge,zhang1,zhang2,stefan1,stefan2,edgexu1,edgexu2,maxboundary,maxboundary2,shang1,toldin1,toldin2,maboundary}. It has been understood since long back that there exists an ordinary boundary condition of a $(2+1)d$ Wilson-Fisher QCP (or a $3d$ classical Wilson-Fisher critical point), where the Landau order parameter $\phi$ has a scaling dimension $\Delta^b_\phi > 1$, which is far greater than the bulk scaling dimension of the order parameter (which is slightly greater than $1/2$). Only recently it has become clear that at the boundary of an $\O(N)$ Wilson-Fisher critical point, in addition to the well-known ordinary boundary criticality, there is a so-called ``{\it extraordinary-log}" boundary criticality, meaning the correlation function of the order parameter at the boundary reads \beqn \langle \phi(\vect{0}) \phi(\vect{x}) \rangle \sim \frac{1}{\left( \ln|\vect{x}| \right)^q}, \eeqn where $q$ depends on $N$. This peculiar scaling was proposed theoretically~\cite{maxboundary,maxboundary2} and recently confirmed numerically in Monte Carlo simulations~\cite{toldin1,toldin2}.

Our current work will bridge these two directions that are under active studying, and we will demonstrate that a $(2+1)d$ QCP under decoherence or weak measurements naturally exhibits the peculiar boundary criticality studied recently. This work is organized as follows:

In section~\ref{general}, we develop the general formalism of
analyzing decohered quantum critical states, including the general connection between the decohered or weakly measured bulk state and the boundary/defect criticality.

In section~\ref{WF} we discuss the $(2+1)d$ $\O(N)$ Wilson-Fisher quantum critical points under decoherence or weak measurements, and in section~\ref{WFlinear} we focus on the quantities linear with the density matrix, which can be observed directly in experiments. We demonstrate that weak measurements generally render the observed quantities rather different from the bulk QCP (with certain postselection that preserves the $\O(N)$ symmetry in the mixed state ensemble), and we may observe the exotic extraordinary-log correlation mentioned above.

In section~\ref{WFnonlinear} we discuss quantities nonlinear with the density matrix, which reveals a lot more structures of the mixed state density matrix of the critical state under decoherence. In particular, we discuss a quantum information phase transition that can be diagnosed through the 2nd Renyi entropy of the decohered system, along with other correlations defined in this section.

Here we would like to point out an important difference between the physical scenarios to be considered in section~\ref{WFlinear} and ~\ref{WFnonlinear}. In section~\ref{WFlinear} we discuss physics under weak measurements on quantities such as energy density, and we allow a postselection on the measurement outcomes. Section~\ref{WFnonlinear} considers quantities nonlinear with the density matrix where postselection is not needed in this scenario; hence it can genuinely correspond to physics under decoherence due to coupling to the environment. 

In section~\ref{lattice} we demonstrate that there is an explicit lattice model with an information transition driven by the strength of decoherence (or weak measurement) analogous to the decoherence-driven transition in section~\ref{WFnonlinear}. We also show that the transition in this lattice model is dual to an information transition in the toric code model, which is related to the error threshold of the toric code~\cite{kitaevpreskill}. 

If we forbid postselection, local correlation functions of the (locally) decohered density matrix would remain largely unchanged from the undecohered pure state. In recent years the nonlocal disorder operator has become a very important diagnosis of quantum states of matter. In section~\ref{nonlocal} we demonstrate that the nonlocal disorder operator can still have qualitatively different behavior from the undecohered pure state density matrix, even if there is no postselection. In particular, we show that in several examples, decoherence can lead to {\it confinement}.

\begin{center}
\begin{figure}
\includegraphics[width=0.48\textwidth]{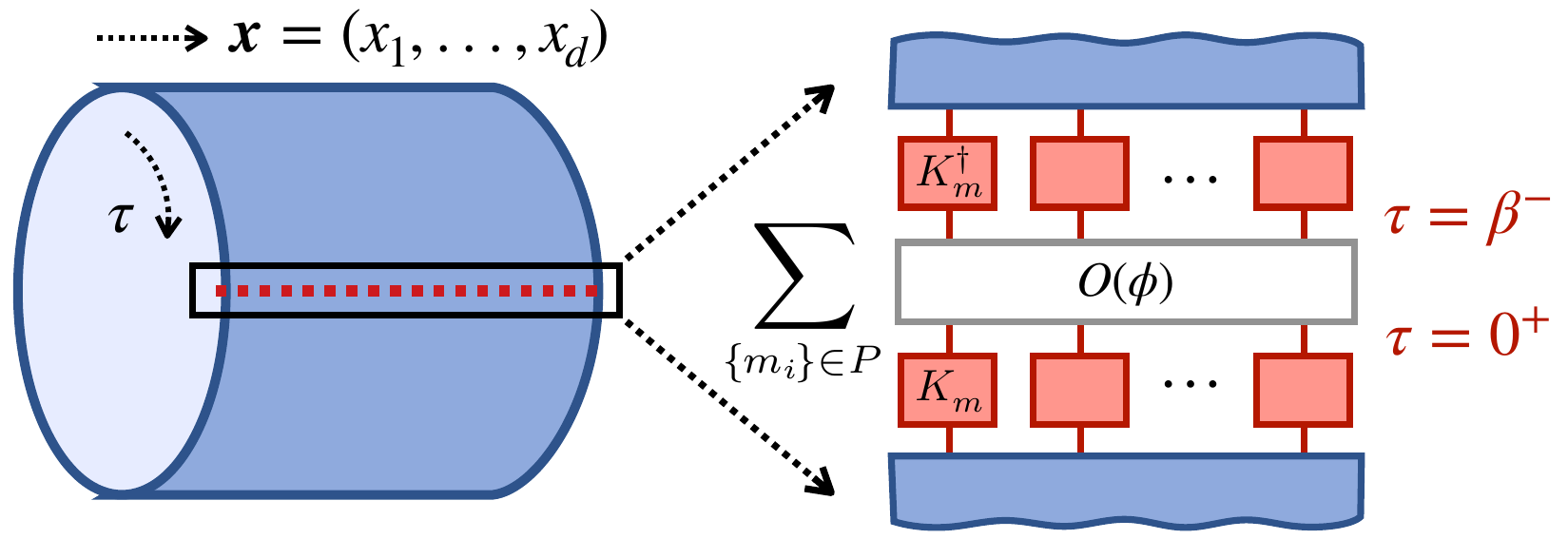}
\vspace{1pt}
\caption{ \textbf{Euclidean spacetime diagram.} The expectation value of operators amounts to evaluating the path-integral in the imaginary space-time with a defect inserted at the $d$-dimensional slab between $\tau=0^+$ and $\tau=\beta^{-}$. Microscopically, the defect (non-trivial $S^{\textrm{int}}$) is captured by the summation over Kraus operators for a local decoherence channel. No postselection corresponds to $P$ being all possible labels of $\{ m_i \}$. Since $\sum_m K_m^\dagger K_m^\vdagger = 1$, if the inserted operator $O(\phi)$ is the product of local operators, its expectation value (correlation function) would only acquire a constant correction. With postselection, the set $P$ is constrained, and the effect of the summation of $P$ corresponds to a non-trivial defect inserted at $\tau = 0$ (or equivalently $\tau = \beta$), and connections to recently studied boundary/defect criticality of $(2+1)d$ QCP can be made. }
\label{spacetime}
\end{figure}
\end{center}

\section{general formalism}
\label{general}

In order to discuss quantum states under decoherence, one approach is to first explicitly derive the ground-state wave function $|\Psi\rangle$ in either exactly soluble lattice models or effective field theories, then construct the density matrix $\hrho^D$ under decoherence~\cite{sptdecohere}. This approach relies on an explicit derivation of the ground-state wave function. The ground-state wave function can be derived for gapped states such as SPT phases~\cite{nishimorilee,sptwf,YouXu2013}, and also gapless phases with a Gaussian (free boson) Lagrangian, such as the $(1+1)d$ Luttinger liquid~\cite{altman}, or the Rokhsar-Kivelson (RK) point in $(2+1)d$ models~\cite{ardonneRK}. But for general interacting theories, such as systems near the Wilson-Fisher quantum critical points, deriving the ground-state wave function is cumbersome. In this section, we follow a more general procedure to study interacting systems under decoherence.

Let us prepare an interacting quantum state which is the ground state of a Hamiltonian, whose pure state density matrix is given as $\hat{\rho}_0$. After the quantum state is prepared, we turn off the Hamiltonian and expose the system to \emph{local} decoherence. For example, for a lattice model of qubits, a decohered density matrix may be represented as 
\begin{align}\label{decohere0}
    & \hrho^D = \mathcal{E}[\hrho_0], \ \ \mathcal{E} = \prod_{\vect{x}} \cE_{\vect{x}}, \nonumber \\ &  \cE_{\vect{x}}[\hrho_0] = (1 - p) \hrho_0 + p Z_{\vect{x}} \hrho_0 Z_{\vect{x}}. 
\end{align}
where $\hat{\rho}^D$ describes a mixed state (ensemble) and $\cE$ is given as the composition of local decoherence channels $\cE_\vect{x}$. One interpretation of this decoherence channel is that, there is a certain probability for the environment to measure qubits in the $Z$ basis (weak measurements) at each location $\vect{x}$, and the measurement outcomes are ``lost", which amounts to the dephasing noise in the study of quantum circuits. A generic decoherence channel maps a density matrix $\hat{\rho}_0$ into $\hat{\rho}^D = \sum_m K_m \hat{\rho}_0  K^\dagger_m$, where $\{K_m\}$ is a set of Kraus operators satisfying the condition $\sum_m K^\dag_m K_m = \openone$. 
If we interpret a decoherence channel to be induced by weak measurements, we can consider a postselection followed by the channel where the resulting density matrix is given as 
\begin{align} \label{eq:postselection}
    \hat{\rho}^D_P \equiv \frac{\cP[\hat{\rho}^D]}{\tr\{ \cP [\hat{\rho}^D] \} },\quad \cP[\hat{\rho}^D] \equiv \sum_{m \in P} K_m \hat{\rho}_0 K^\dagger_m \end{align}
where $\cP$ is a generalized projection onto a subset of measurement outcomes $P$, i.e., postselection. 

In this work, we will mostly study systems at or close to a QCP, hence we will use a coarse-grained continuous space rather than a lattice model. In the coarse-grained formalism, the density matrix of a pure state is given by the following imaginary time path-integral  \begin{align}
    &[\hrho_0]_{\phi_1(\vect{x}), \phi_2(\vect{x})} = \langle \phi_1(\vect{x}) | \Psi \rangle \langle \Psi| \phi_2(\vect{x}) \rangle  \nonumber \\
    &\qquad \sim  \lim_{\beta \ra \infty } \int_{\substack{ \phi(\vect{x}, 0) = \phi_1(\vect{x}) \\ \phi(\vect{x},  \beta) = \phi_2(\vect{x}) } } D \phi(\vect{x}, \tau) \exp(- \cS), 
\end{align} 
where $\cS = \int_0^\beta d\tau d\vect{x} \ \cL (\phi) $ is the bulk action of the system and $\vect{x}=(x_1,...,x_d)$ is the spatial coordinate. Following Ref.~\onlinecite{altman,sptdecohere}, a class of decoherence problem in the coarse-grained continuous space can be converted into the following imaginary-time path-integral:  
\begin{widetext} \beqn && [\hrho^D]_{\phi_1(\vect{x}), \phi_2(\vect{x})} \sim \lim_{\beta \ra \infty } \int_{\substack{ \phi(\vect{x}, 0) = \phi_1(\vect{x}) \\ \phi(\vect{x},  \beta) = \phi_2(\vect{x}) } } D \phi(\vect{x}, \tau) \exp \left( - \cS - \cS^{\Int} \right); \cr\cr\cr && \cS = \int_0^\beta d\tau d\vect{x} \ \cL (\phi), \ \ \ \cS^{\Int} = \int d\vect{x} \ \cL^{\Int} ( \phi(\vect{x}, 0), \phi(\vect{x}, \beta) ). \label{decohere1} \eeqn \end{widetext}
The effect of decoherence or postselection is captured by an extra interaction term $\cL^{\Int}$ in the Lagrangian, and $\cL^{\Int}$ has no temporal integral, i.e., it is a ``long range" interaction in the Euclidean temporal direction, between fields at $\tau = 0$ and $\tau = \beta$. The coupling constant in $\cL^{\Int}$ is controlled by the strength of decoherence (or strength of weak measurement) $p$.  A natural choice of $\cL^{\Int}$ favors configurations with $\phi(\vect{x}, 0)\sim \phi(\vect{x}, \beta)$, meaning it enhances the weight of the diagonal components of the density matrix, which drives the system into a mixed state density matrix.

The most important factor that determines the form of $\cL^{\Int}$ is still its symmetry. Let us label the symmetry of the original Lagrangian $\cL$ as $G$, and assume that the original pure state without decoherence is a symmetric state under $G$. Then there are two types of symmetry constraints on $\cL^{\Int}$. If the environment is weakly ``measuring" quantities that are invariant under $G$, then $\cL^{\Int}$ must be invariant under a ``doubled" symmetry transformation, i.e., it is invariant under symmetry transformation $G$ on $\phi(\vect{x}, 0)$ and $\phi(\vect{x}, \beta)$ separately. In the language of Ref.~\onlinecite{sptdecohere}, this is the case that preserves the doubled $G^u \times G^l$ symmetry, where $G^u$ and $G^l$ correspond to the upper and lower symmetry in the formalism of doubled Hilbert space using the Choi-Jamiolkowski isomorphism~\cite{JAMIOLKOWSKI1972,CHOI1975}, which maps any density matrix to a pure ket-state in the doubled Hilbert space. However, if the environment is measuring quantities that carry a nontrivial representation of $G$, but eventually we sum over the measurement outcomes within the same representation of $G$ with equal weight, meaning the symmetry $G$ is broken in each quantum trajectory but still preserved in an average sense, then $\cL^{\Int}$ is only invariant under symmetry $G$, which is the diagonal subgroup of $G^u \times G^l$, and it corresponds to a simultaneous transformation of $\phi(\vect{x}, 0)$ and $\phi(\vect{x}, \beta)$.

In \eqnref{decohere0}, if the original pure state is invariant under a $\mathbb{Z}_2$ symmetry action $\prod_\vect{x} X_\vect{x}$ 
then the pure state density matrix $\hrho_0$ is invariant under a doubled $\mathbb{Z}_2^u \times \mathbb{Z}_2^l$ symmetry, where $\mathbb{Z}_2^u$ and $\mathbb{Z}_2^l$ correspond to the left and right operations of the $\mathbb{Z}_2$ symmetry. But since the decoherence is caused by weakly measuring the Ising variable which carries a nontrivial representation of $\mathbb{Z}_2$, the decohered density matrix $\hrho^D$ is no longer invariant under separate $\mathbb{Z}_2^u$ or $\mathbb{Z}_2^l$ action, it is instead only invariant under the simultaneous action of both $\mathbb{Z}_2^u$ and $\mathbb{Z}_2^l$, hence $\hrho^D$ is invariant under a diagonal $\mathbb{Z}_2$ symmetry. Here the $\mathbb{Z}_2^u \times \mathbb{Z}_2^l$ symmetry is {\it explicitly} broken down to $\mathbb{Z}_2$ by the decoherence; later we will also discuss an example where the decoherence preserves the doubled symmetry but the doubled symmetry can be {\it spontaneously} broken down to the diagonal $\mathbb{Z}_2$.

Now suppose we would like to compute the expectation value of a certain quantity $O(\hphi)$ that is a composite of $\hphi$. The expectation value is linear with the density matrix, and it can be evaluated in the path integral form: 
\beqn \tr\{ \hrho^D O(\hphi) \} \sim \int_{\phi(\vect{x}, 0) = \phi(\vect{x}, \beta)} D\phi(\vect{x}, \tau) \cr\cr \times O (\phi_{\tau=0}) \exp\left( - \cS - \cS^{\Int} \right),
\label{linear}
\eeqn 
where we have identified the field configurations $\phi(\vect{x}, 0)$ and $\phi(\vect{x}, \beta)$ due to the trace.  If there is no postselection at all, one can easily show that $\tr\{\hrho^D O(\hphi)\}$ and $\tr\{\hrho_0 O(\hphi)\}$ have essentially the same behavior (except for some local corrections) as long as $O(\hphi)$ is a product of local operators~\footnote{Note that $\tr\{ \cE[\rho] O \} = \tr\{ \rho \cE^*[O] \}$, where if $\cE[\rho] = \sum_i K_i \rho K_i^\dagger$ then $\cE^*[O] = \sum_i K^\dagger_i \rho K_i$. Accordingly, the correlation function only acquires a local correction. }, as pictorially illustrated in \figref{spacetime}. For example, the correlation function $\tr\{\hrho^D \hphi(\vect{x}_1) \hphi(\vect{x}_2)\}$ should have the same scaling in space as $\tr\{\hrho_0 \hphi(\vect{x}_1) \hphi(\vect{x}_2)\}$. In the field theory language, this corresponds to $\cS^{\textrm{int}}$ being trivial when $\phi(\vect{x},0) = \phi(\vect{x},\beta)$.

But if there is some weak postselection by $\cP$ even on quantities that are singlet under $G$, $\tr\{\hrho^D_P \hphi(\vect{x}_1) \hphi(\vect{x}_2)\}$ can be very different from $\tr\{\hrho_0 \hphi(\vect{x}_1) \hphi(\vect{x}_2)\}$. With postselection, $\cS^{\Int}$ remains non-trivial even at $\phi(\vect{x},0) = \phi(\vect{x},\beta)$, which effectively corresponds to the insertion of defects at the slab $\tau = 0$ (or $\tau = \beta$) in the path-integral in the $(d+1)-$dimensional Euclidean space-time. This is where we can make a connection to all the recent studies on boundary criticality for systems with $d=2$~\cite{max3}, and the desired decohered correlation functions become the correlation functions restricted on the plane-like defect.

\begin{center}
\begin{figure}
\includegraphics[width=0.47\textwidth]{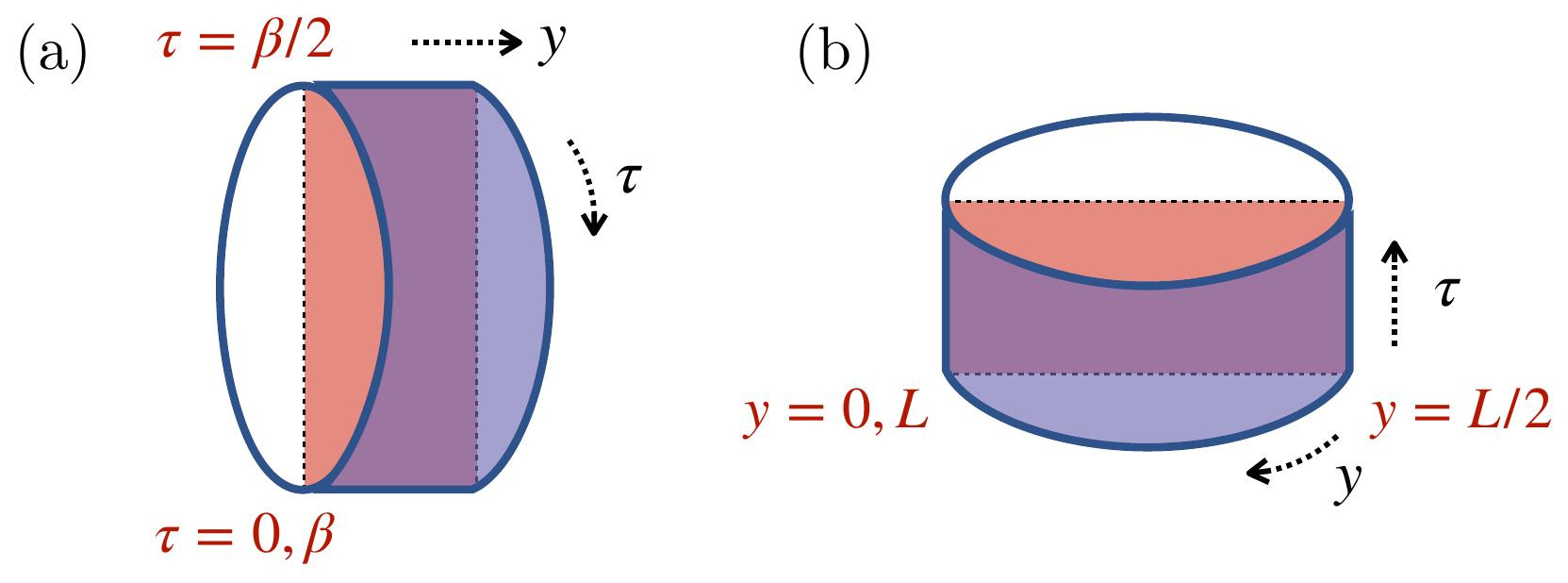}
\caption{ \textbf{Euclidean spacetime diagram and its space-time rotated version.} The evaluation of the second Renyi entropy of $\hrho^D$ becomes a path-integral in the Euclidean space-time with an interaction between fields at $\tau = 0$ and $\tau = \beta/2$. In the space-time rotated picture, such an interaction corresponds to the long-range interaction between fields at $y = 0$ and $y = L / 2 $, where $L=\beta$. Note that the first coordinate of $\vect{x}=(x,y)$ is not shown in the diagram.} \label{spacetime1}
\end{figure}
\end{center}

The expectation value of an operator is linear with the density matrix, which is directly related to experimental observables. But a lot of information of the quantum system is encoded in quantities that are {\it nonlinear} with the density matrix. The most famous example of such is the von Neumann entropy of the density matrix. In this work, we will evaluate quantities such as the 2nd Renyi entropy of $\hrho^D$, which is an analogue of the von Neumann entropy, and it also provides an approximate characterization~\footnote{Formally, we can evaluate the $n$-th Renyi entropy, and takes the limit $n\rightarrow 1$ to obtain the behavior of the von Neumann entropy.} of the amount of quantum information lost due to entangling with the environment: 
\begin{align} \label{renyi}
    S^{(2)} &= - \log \tr\{(\hrho^D)^2\} \nonumber \\ &\sim  - \log \lim_{ \beta \ra \infty} \int D \phi(\vect{x}, \tau)\, \times \nonumber \\ & \,\,\exp\left( - \cS - \int d\vect{x} \ \cL^{\Int} ( \phi(\vect{x}, 0), \phi(\vect{x}, \beta/2) ) \right). 
\end{align}
This calculation is schematically shown in Fig.~\ref{spacetime1}, which amounts to evaluating the partition function and free energy of the system with an interaction between fields at imaginary time $\tau = 0$ and $\tau = \beta/2$. Or we can also rotate the space-time (assuming there is a Lorentz symmetry), then the problem becomes evaluating the partition function of the system with nonlocal interaction in space, but constant in time.  

Other quantities of interest include $\tr\{(\hrho^D)^2 O(\hphi)\}$ or $\tr\{\hrho^D O(\hphi)\hrho^D O(\hphi) \}$. We will show that these quantities nonlinear with $\hrho^D$ would reveal some novel quantum phase transitions. In the example we will discuss in the next section, the decoherence respects the doubled $\mathbb{Z}_2^u \times \mathbb{Z}_2^l$ symmetry; but when we increase the decoherence strength, the 2nd Renyi entropy of the system (which captures the information loss to the environment) may encounter singularity at a critical decoherence strength, which corresponds to the {\it spontaneous} symmetry breaking from $\mathbb{Z}_2^u \times \mathbb{Z}_2^l$ to the diagonal $\mathbb{Z}_2$.

\section{Decohered Wilson-Fisher critical point} \label{WF}

In this section, we study two different scenarios for the system under weak measurements: In \secref{WFlinear}, we postselect a set of measurement outcomes and show that the resulting correlation functions linear in the density matrix $\hrho^D_P$ exhibit novel behaviors. In \secref{WFnonlinear}, we average over all measurement outcomes, which corresponds to genuine decoherence; although correlation functions linear in $\hat{\rho}^D$ would exhibit an ordinary Wilson-Fisher behavior in this case, we show that quantities non-linear in $\hat{\rho}^D$ still exhibit novel behaviors including the extraordinary-log criticality and information-theoretic phase transition.

\subsection{Quantities linear with \texorpdfstring{$\hrho^D_P$}{rhoDP}} \label{WFlinear}

First, we discuss critical physics under postselection as elaborated in \eqnref{eq:postselection}. The critical system we consider is the O($N$) Wilson-Fisher fixed point in $(2+1)d$, where $\vect{x}=(x,y)$ is a two-dimensional spatial coordinate.
Under postselection onto a set of measurement outcomes $P$, we would like to consider quantities linear with $\hrho^D_P$, for example the correlation function $\tr\{\hrho^D_P \hat{\vect{\phi}}(0) \cdot \hat{\vect{\phi}}(\vect{x})\}$. The evaluation of quantities such as $\tr\{ \hrho^D_P O(\hat{\vect{\phi}}(\vect{x}))\}$ can be performed by evaluating path-integral \eqnref{linear}. Since we would like to keep at least the diagonal $\O(N)$ symmetry, the postselected measurement outcomes $P$ should be invariant under $\O(N)$ symmetry. Under that condition, the corresponding $\cS^{\Int}$ becomes a function of the magnitude of the order parameter $|\vect{\phi}|$ at $\tau = 0$, which is always equal to $|\vect{\phi}|$ at $\tau = \beta$ due to the trace. The simplest term (and most relevant term) of $\cS^{\Int}$ is \beqn \cS^{\Int} = \int dxdy \ \varepsilon |\vect{\phi}(\vect{x},0)|^2. \eeqn The term $\cS^{\Int}$ can be interpreted as weakly measuring the energy density or the order parameter $\vect{\phi}$ of the system, followed by postselection.

Although there is only one simple term in $\cS^{\Int}$, the physical consequence is already rather nontrivial. $\cS^{\Int}$ is a $2d$ interface in a $(2+1)d$ cylindrical space-time with extra mass $\varepsilon$ for the order parameter, and obviously $\varepsilon$ is always a relevant perturbation, as the scaling dimension of $|\vect{\phi}|^2$ is always smaller than 2 at the $\O(N)$ Wilson-Fisher fixed point. When $\varepsilon > 0$, $\cS^{\Int}$ suppresses the fields at $\tau = 0$, and ``cuts" the connection between $\tau = 0^+$ and $\tau = \beta^-$, i.e. $\tau = 0^+$ and $\tau = \beta^-$ become two boundaries with ordinary boundary condition. In this case, the correlation function $\tr\{ \hrho^D_P \ \hat{\vect{\phi}}(\vect{0}) \cdot \hat{\vect{\phi}}(\vect{x}) \}$ scales as~\cite{boundary1n} 
\beqn \tr\{ \hrho^D_P \ \hat{\vect{\phi}}(\vect{0}) \cdot \hat{\vect{\phi}}(\vect{x}) \} \sim \frac{1}{|\vect{x}|^{2\Delta^b_\vect{\phi}}}, \eeqn where \beqn \Delta^b_{\vect{\phi}} = 1 + \frac{2}{3N} + O\left( \frac{1}{N^2} \right) \eeqn is the boundary scaling dimension of the order parameter $\vect{\phi}$ to the first order expansion of $1/N$. Note that the value of $\Delta^b_{\vect{\phi}}$ is far larger than the bulk scaling dimension of $\vect{\phi}$.

When $\varepsilon < 0$, the latest progress~\cite{maxboundary2, max3} of the boundary criticality indicates that $\cS^{\Int}$ will drive the interface $\tau = 0$ into an extraordinary-log criticality, which implies that the correlation function becomes \beqn \tr\{ \hrho^D_P \, \hat{\vect{\phi}}(\vect{0}) \cdot \hat{\vect{\phi}}(\vect{x}) \} \sim \frac{1}{(\ln|\vect{x}|)^q}. \eeqn Please note that, for a single exposed boundary against the vacuum, the extraordinary-log criticality exists only for $N < N_c$, with the critical $N_c$ estimated around $N_c \sim 5$~\cite{maxboundary2}; but for an interface defect inserted in space-time, the theoretical prediction is that $N_c \rightarrow \infty$~\cite{max3}. 

\begin{figure}
\includegraphics[width=0.39\textwidth]{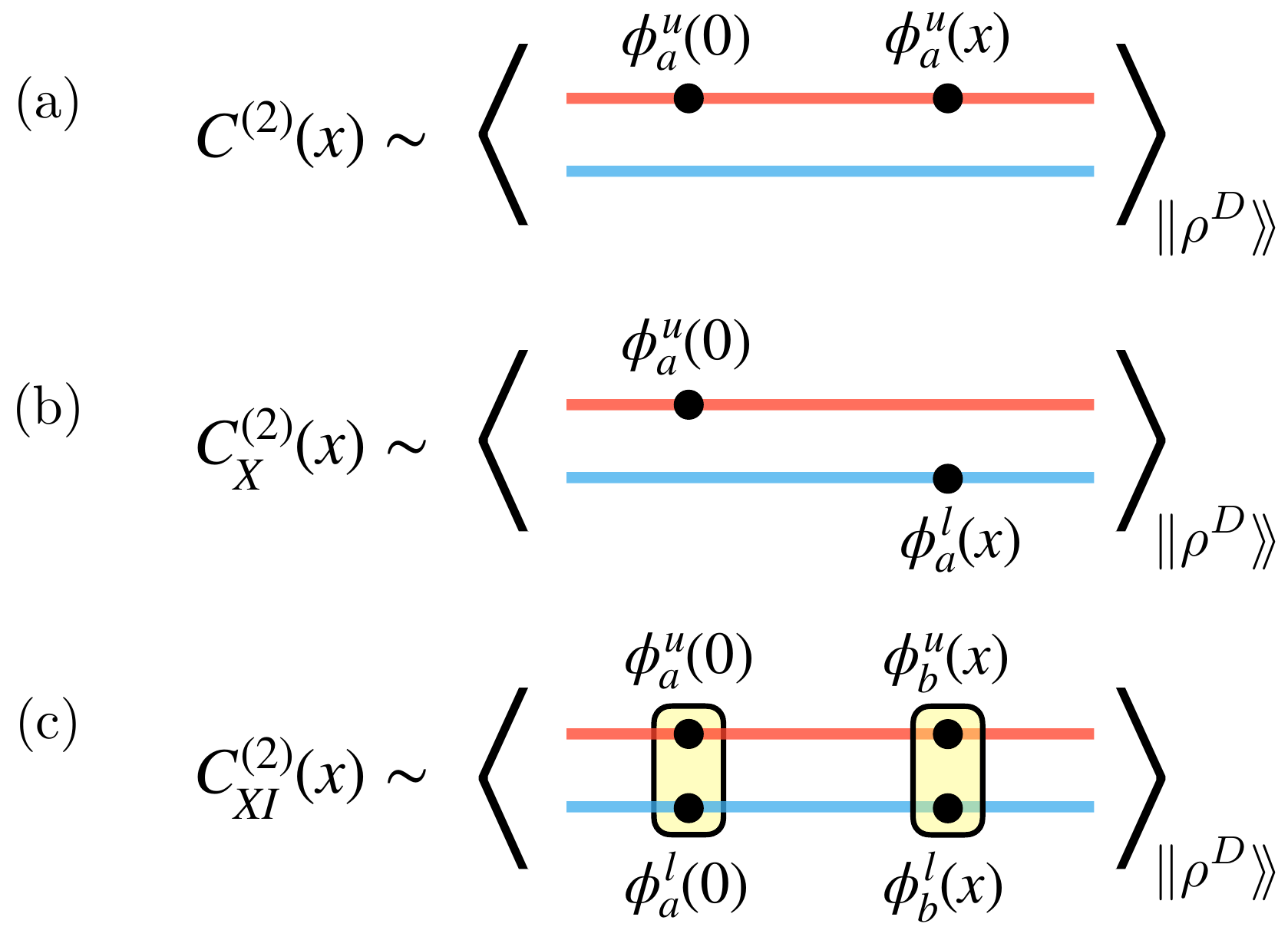}
\caption{ \textbf{Correlation functions in the doubled Hilbert space.} The red and blue lines represent the upper and lower Hilbert spaces respectively in the doubled Hilbert space (See \secref{sec:choi}).} 
\label{fig:corrs}
\end{figure}

\subsection{Quantities nonlinear with \texorpdfstring{$\hrho^D_P$}{rhoDP}} \label{WFnonlinear}

Now we evaluate quantities nonlinear with $\hrho^D$. Although our formalism can be straightforwardly generalized for any higher orders of $\hrho^D$, here we focus on the quantities that are quadratic in $\hrho^D$. The quantities of interest include the 2nd Renyi entropy $S^{(2)}$, the correlation function $C^{(2)}(\vect{x})$, the ``crossed" correlation function $C^{(2)}_X(\vect{x})$, and the ``crossed-Ising" correlation function $C^{(2)}_{XI}(\vect{x})$ defined as follows: 
\begin{align}\label{eq:corrs}
    S^{(2)} =& - \log \tr\{ (\hrho^D)^2 \}, \nonumber \\ C^{(2)}(\vect{x}) \sim& \,\, \tr\{ (\hrho^D)^2 \hat{\vect{\phi}}(\vect{0}) \cdot \hat{\vect{\phi}}(\vect{x})\}, \nonumber \\ C^{(2)}_X(\vect{x}) \sim& \sum_a \tr\{ \hrho^D \hat{\phi}_a (\vect{0}) \hrho^D \hat{\phi}_a (\vect{x})\} \nonumber \\ 
    C^{(2)}_{XI}(\vect{x}) \sim& \sum_{a \neq b} \tr\{ \hrho^D \hat{\phi}_a (\vect{0}) \hat{\phi}_b (\vect{x}) \hrho^D \hat{\phi}_a (\vect{0}) \hat{\phi}_b (\vect{x})\}.
\end{align}
The expression of the correlation functions above need to be divided by the \emph{purity} of the decohered density matrix $\tr\{(\rho^D)^2\}$ to be properly normalized. We note that these correlation functions are distinguished by their representation under $\O(N)^u \times \O(N)^l$ symmetry, which becomes clear in the doubled Hilbert space as in \figref{fig:corrs}.

As an example, we consider the following interaction term $\cS^{\Int}$ in Eq.~\ref{renyi}: \begin{align}
    \cS^{\Int} &= \int dxdy \ \Big[ W \left( |\vect{\phi}(\vect{x},0)|, \ |\vect{\phi}(\vect{x},\beta/2)| \right) \nonumber \\
    &\qquad -\, w \left( \vect{\phi}(\vect{x}, 0) \cdot \vect{\phi} (\vect{x}, \beta/2) \right)^2 \Big]. \label{Sint1}
\end{align}
Here $W$ is still a function of the magnitude of the order parameter at $\tau= 0$ and $\tau = \beta/2$. 
The two terms in Eq.~\ref{Sint1} correspond to two different types of weak measurements (decoherence) channels. 
The first term still corresponds to weakly measuring the energy density of the system, which preserves the doubled symmetry of the system, resulting in the coupling
\begin{align}
&    \int dxdy \ W \left(
|\vect{\phi}(\vect{x},0)|, \ |\vect{\phi}(\vect{x},\beta/2)| \right) \nonumber \\ & ~~~~~~ = \int dxdy \ \frac{\varepsilon}{2}(|\vect{\phi}(\vect{x},0)|^2 + |\vect{\phi}(\vect{x},\beta/2)|^2) + ... \end{align} where the ``..." part includes quartic and higher-order terms in $|\vect{\phi}(\vect{x},0)|$ and $|\vect{\phi}(\vect{x},\beta/2)|$, such as $|\vect{\phi}(\vect{x},0)|^2|\vect{\phi}(\vect{x},\beta/2)|^2$. The second $w$ term can be rewritten as 
\begin{align}
  &-w \left( \vect{\phi}(\vect{x}, 0) \cdot \vect{\phi} (\vect{x}, \beta/2) \right)^2 \sim \nonumber \\ &\qquad - w \sum_{a,b = 1}^N \left( Q_{ab}(\vect{x}, 0) Q_{ab}(\vect{x}, \beta/2) \right) + \cdots,
\end{align}
where $Q_{ab} = \phi^a\phi^b - \frac{1}{N}\delta_{ab}|\vect{\phi}|^2$ is the traceless rank-2 symmetric tensor of the O($N$) vector $\vect{\phi}$, and the ellipsis are terms that only depend on $|\vect{\phi}|$ and can be absorbed into $W$. The $w$ term can be interpreted as a weak measurement on the tensor $Q_{ab}$, and eventually all the measurement outcomes are summed with equal weight. The $w$ term breaks the $\SO(N)^u \times \SO(N)^l$ down to the diagonal $\SO(N)$, but still preserves the $\mathbb{Z}_2^u \times \mathbb{Z}_2^l$ symmetry, where $\mathbb{Z}_2$ corresponds to changing the sign of $\vect{\phi}$.

One can perform the space-time rotation in the $(y, \tau)$ plane, so the two interfaces at $\tau = 0, \beta/2$ become spatial interfaces at $y = 0$ and $L/2$, with $L = \beta$. The interaction term then becomes \beqn \cS^{\Int} &=& \int d\tau dx \ W \left( |\vect{\phi}(x, \tau)|_{y = 0}, \ |\vect{\phi}(x, \tau)|_{y = L/2} \right) \cr\cr &&- \,w \left( \vect{\phi}(x,\tau)_{y = 0} \cdot \vect{\phi}(x,\tau)_{y = L/2} \right)^2. \label{Sint1wick} \eeqn 
The first term $W$ will either explicitly include an extra mass term $\varepsilon |\vect{\phi}|^2$ at the interface $y = 0$ and $y = L/2$, or generate the mass term through renormalization group flow. Then depending on the sign of $\varepsilon$, there can be three possible scenarios:

(1) If $\varepsilon > 0$, the extra mass term $\varepsilon |\vect{\phi}|^2$ at the interface $y = 0$ and $y = L/2$ is a relevant perturbation. The role of this extra mass term is to ``cut" the system into two halves: the region from $y \in (0, L/2)$ and region $y \in (L/2, L \sim 0)$. The relevant mass term will make the two interfaces at $y = 0$ and $y = L/2$ both at ordinary boundary criticality.

In this scenario, $w$ is obviously an irrelevant perturbation since both interfaces $y = 0$ and $y = L/2$ have ordinary boundary criticality, and the scaling dimension of $\vect{\phi}$ is greater than 1 at the interfaces. Hence the directions of $\vect{\phi}$ at the two interfaces are pretty much uncorrelated to each other. Then we expect the correlation function, the crossed-Ising correlation, and the crossed-correlation functions to behave as \beqn C^{(2)} &\sim& \tr\{ (\hrho^D)^2 \hat{\vect{\phi}}(\vect{0}) \cdot \hat{\vect{\phi}}(\vect{x}) \} \sim \frac{1}{|\vect{x}|^{2\Delta^b_{\vect{\phi}}}}, \cr\cr C^{(2)}_{XI} &\sim& \sum_{a \neq b} \tr\{ \hrho^D \hat{\phi}_a (\vect{0}) \hat{\phi}_b (\vect{x}) \hrho^D \hat{\phi}_a (\vect{0}) \hat{\phi}_b (\vect{x}) \} \sim 0, \cr\cr C^{(2)}_X &\sim& \sum_a \tr\{ \hrho^D \hat{\phi}_a (\vect{0}) \hrho^D \hat{\phi}_a (\vect{x}) \} \sim 0. \eeqn  For example, the crossed-correlation $C^{(2)}_X$ corresponds to the correlation function between order parameters at the two different interfaces; since $w$ is irrelevant when $\varepsilon > 0$, the directions of order parameters at the two interfaces are uncorrelated, hence $C^{(2)}_X$ should vanish in the limit $\beta, L \rightarrow \infty$. Another way to perceive these results is that, since $w$ is irrelevant, the system should have a full $\SO(N)^u \times \SO(N)^l$ symmetry in the infrared, but the correlation function $C^{(2)}_X$ and $C^{(2)}_{XI}$ break the $\SO(N)^u \times \SO(N)^l$ symmetry, hence they must vanish.

(2) In the case with $\varepsilon < 0$, the extra local mass term $\varepsilon$ is still a relevant perturbation, and it flows to the extraordinary-log criticality. Then the $w$ term (we assume $w > 0$) is a very relevant perturbation, and it will flow to $w \rightarrow \infty$.

The fate of the system with strong $w$ can be perceived through a mean field decoupling of $\cL^{\mathrm{int}}$. We first recognize that the $w$ term is analogous to the coupling between two sets of N\'{e}el order parameters in the $J_1-J_2$ Heisenberg model on the square lattice, which has applications in the context of frustrated magnets and iron-pnictides superconductors~\cite{henley,nematiclarkin,ironsi,nematichu,nematicxu}. Guided by the previous studies in these contexts, the most natural mean field decoupling of the $w$ term is \beqn &&- w \left( \vect{\phi}(x,\tau)_{y = 0} \cdot \vect{\phi}(x,\tau)_{y = L/2} \right)^2 \sim \cr\cr &&\quad - 2 w \Phi(x, \tau) \left( \vect{\phi}(x,\tau)_{y = 0} \cdot \vect{\phi}(x,\tau)_{y = L/2} \right) \cr\cr &&\quad + w \Phi(x, \tau)^2. \eeqn Here we have introduced an Ising order parameter $\Phi(x, \tau) \sim \left( \vect{\phi}(x,\tau)_{y = 0} \cdot \vect{\phi}(x,\tau)_{y = L/2} \right)$. The order parameter $\Phi$ is analogous to the nematic order parameter in the $J_1-J_2$ Heisenberg model on the square lattice, and the phase with large $w$ is likely a phase with condensation of $\Phi$, which spontaneously breaks $\mathbb{Z}^u_2 \times \mathbb{Z}^l_2$ down to the diagonal $\mathbb{Z}_2$.

The condensation of $\Phi$ will ``pin" $\hat{\vect{\phi}}_{y = 0}$ and $\hat{\vect{\phi}}_{y = L/2}$ along the parallel direction. With a nonzero condensate of $\Phi$, the three correlation functions mentioned above behave like  
\beqn \label{eq:crossed_Ising}  
C^{(2)} &\sim& \tr\{ (\hrho^D)^2 \hat{\vect{\phi}}(\vect{0}) \cdot \hat{\vect{\phi}}(\vect{x})\} \sim \frac{1}{(\ln |\vect{x}|)^q}, \cr\cr 
C^{(2)}_{XI} &\sim& \sum_{a \neq b} \tr\{ \hrho^D \hat{\phi}_a (\vect{0}) \hat{\phi}_b (\vect{x}) \hrho^D \hat{\phi}_a (\vect{0}) \hat{\phi}_b (\vect{x}) \} 
\cr\cr &\sim& \mathrm{Const}, \cr\cr 
C^{(2)}_X &\sim& \sum_a \tr\{ \hrho^D \hat{\phi}_a (\vect{0}) \hrho^D \hat{\phi}_a (\vect{x})\} \sim \frac{1}{(\ln |\vect{x}|)^q}.  \eeqn  The crossed-Ising correlation saturates to a nonzero constant in the limit of large $|\vect{x}|$ due to the condensate of $\Phi$, which also leads to an extraordinary-log correlation of crossed-correlation function. When $w$ flows to infinity and $\Phi$ condenses, there is only one $\O(N)$ symmetry in the infrared; hence all these correlations can be nonzero.

(3) Now if $\varepsilon = 0$, with large but finite $N$, the scaling dimension of the traceless rank-2 symmetric tensor $Q_{ab}$ is $\Delta = 1 + 32/(3\pi^2 N)$ to the leading order of $1/N$ expansion. This means that $w$ is weakly irrelevant with scaling dimension $[w] = - 64/(3\pi^2 N)$. Then the beta function of $w$ should be \beqn \beta(w) = \frac{dw}{d\ln l} = - \frac{A}{N}w + C w^2, \eeqn where $A = 64/(3\pi^2)$. The constant $C$ can be extracted through some OPE calculation, or simply a one-loop calculation in the large$-N$ limit. Since $C$ is positive, there is a fixed point at finite $w_\ast$, beyond which $w$ will flow strongly, and the order parameter $\Phi$ defined above will condense.

\subsection{The phase diagram}

\begin{figure}
\includegraphics[width=0.37\textwidth]{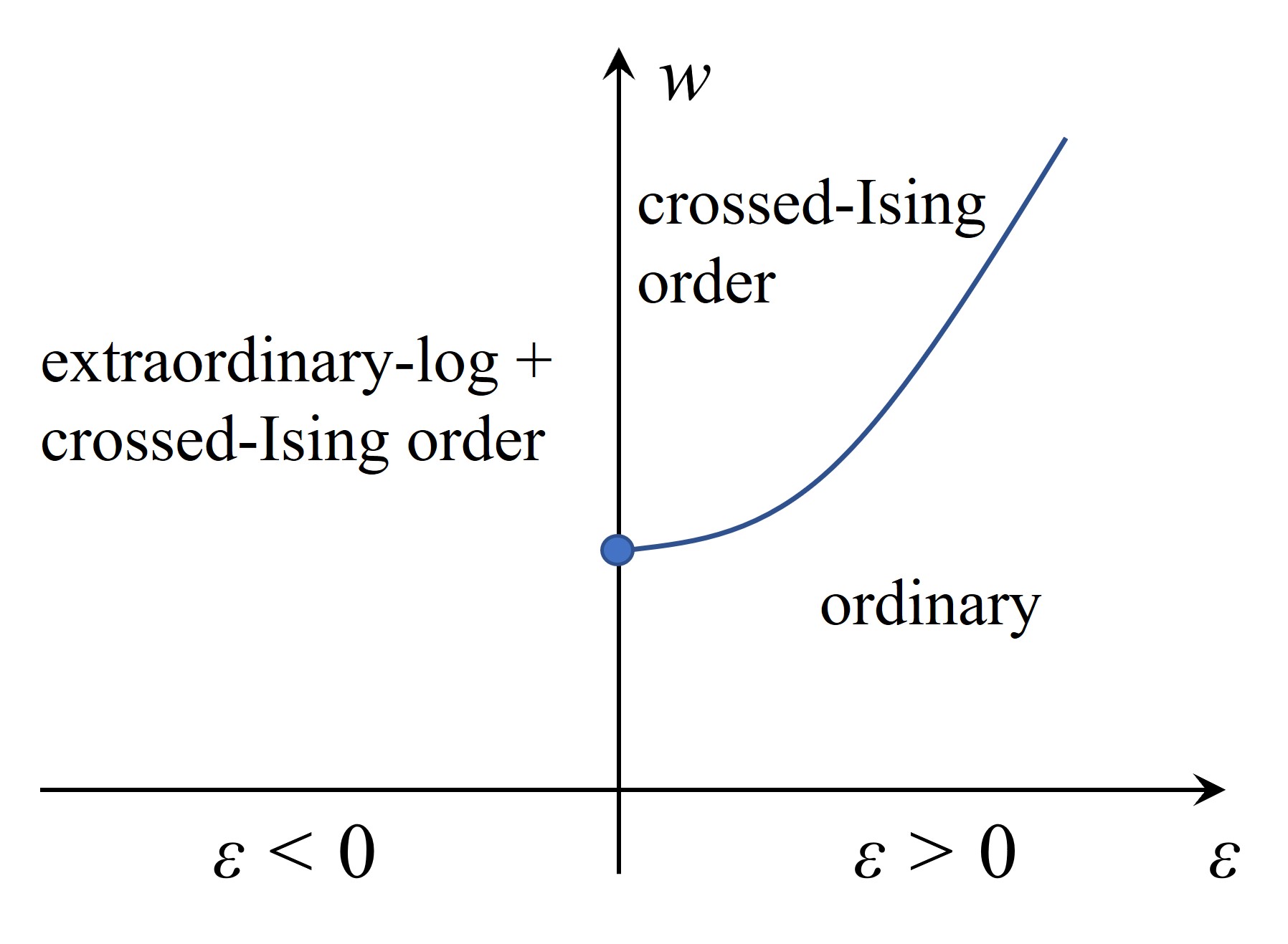}
\caption{ \textbf{Phase Diagram.} The global phase diagram of decohered Wilson-Fisher critical point in terms of quantities nonlinear in $\hrho^D$.} \label{phasedia}
\end{figure}

In phase diagram Fig.~\ref{phasedia} we summarize the results related to $(\hrho^D)^2$ discussed in this section. The order-disorder transition of $\Phi$ should extend to the region with $\varepsilon > 0$, where there is a competition between $\varepsilon$ which drives the interfaces to the ordinary boundary criticality and $w$ which drives the crossed-Ising transition. In this phase diagram, the critical $w_c$ is a function of $\varepsilon$: $w_c(\varepsilon) - w_c(0) \sim \varepsilon^{\Delta_{w}/\Delta_{\varepsilon}}$, and $\Delta_{w,\varepsilon}$ are the scaling dimensions of parameter $w$ and $\varepsilon$.

Since $\Phi(\vect{x}) \sim \vect{\phi}(\vect{x}, 0)\cdot \vect{\phi}(\vect{x}, \beta/2)$ is the crossed-Ising order parameter, the crossed-Ising correlation function should show a transition from short-range to  correlation while increasing $w$. If we consider $\tr\{ (\hrho^D)^2 \}$ as the partition function, and the 2nd Renyi entropy $S^{(2)} = - \log \tr\{ (\hrho^D)^2 \}$ as the free energy, the nature of the transition at $w = w_c$ and $\varepsilon > 0$ should belong to a $2d$ classical Ising universality class. Without coupling to other modes with nonlocal correlations in the infrared, the order-disorder transition of an Ising order parameter in the $2d$ space should belong to the $2d$ Ising universality class, and here we should consider the perturbation of the coupling $w$ on top of the $2d$ Ising transition. In fact, if we start with a $2d$ classical Ising transition of order parameter $\Phi$, the coupling $- w \Phi(\vect{x}) \left( \vect{\phi}(\vect{x}, 0)\cdot \vect{\phi}(\vect{x}, \beta/2) \right)$ is irrelevant knowing the fact that the scaling dimension $\Delta^b_{\vect{\phi}} > 1$ for the ordinary boundary criticality at $\varepsilon > 0$. Hence at $w = w_c$ and $\varepsilon > 0$, the crossed-Ising correlation function should scale as \beqn C^{(2)}_{XI}(\vect{r}) \sim \frac{1}{|\vect{r}|^{1/4}}. \eeqn Also, when we increase $w$ across $w_c$, the 2nd Renyi entropy $ S^{(2)} = - \log \tr\{ (\hrho^D)^2\}$ should have the same singularity as the free energy of the classical $2d$ Ising model.

\begin{figure}
\includegraphics[width=0.49\textwidth]{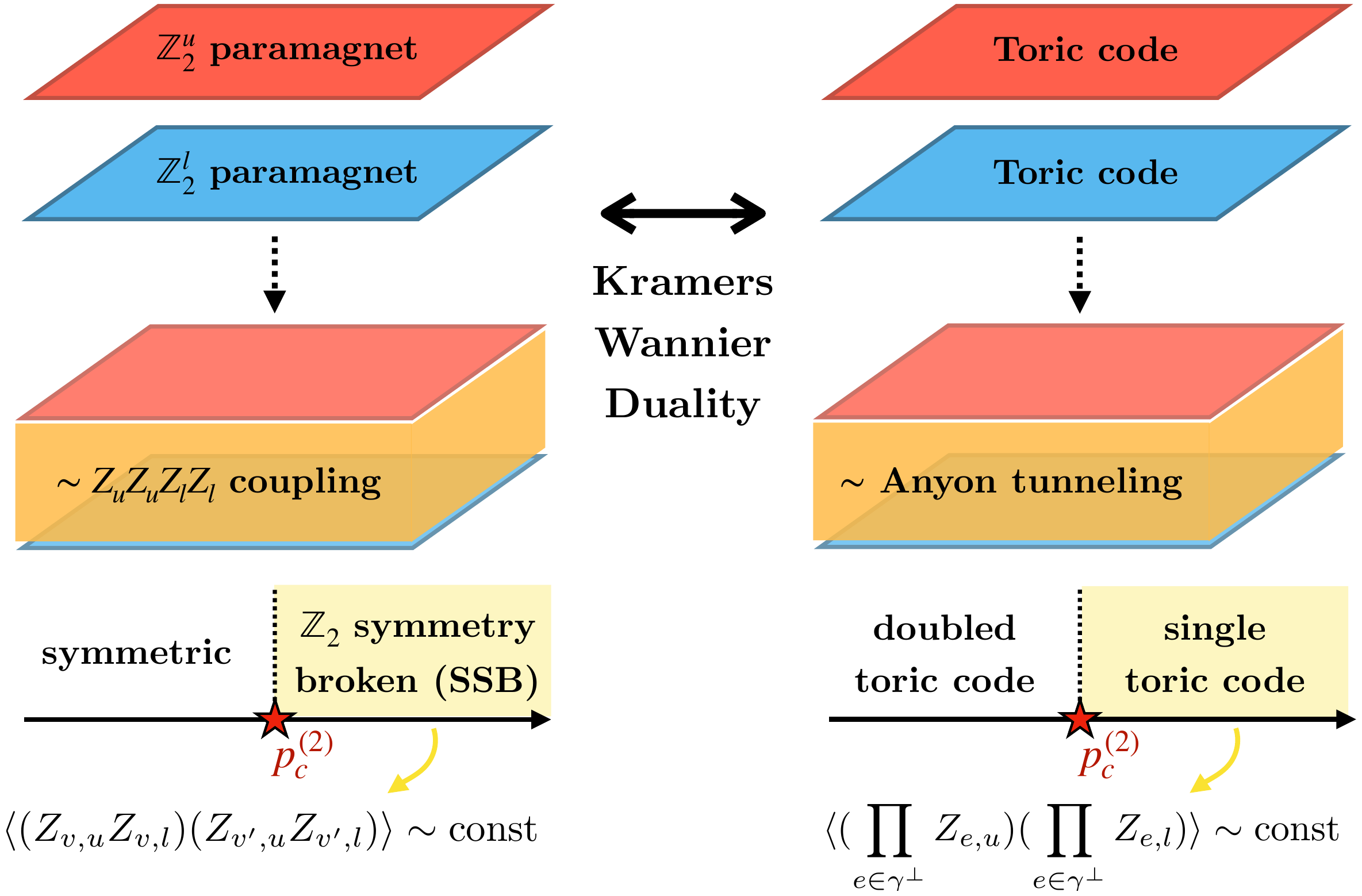}
\caption{ \label{fig:bilayer} \textbf{Choi Isomorphism and Duality.} Schematic diagrams illustrate how decohered mixed states map into the coupled bilayer system under the Choi-Jamiolkowski isomorphism. Through the Kramers-Wannier duality, $\mathbb{Z}_2$ paramagnet under symmetric decoherence (left) maps into the $\mathbb{Z}_2$ toric code under dephasing noise (right). Accordingly, their phase diagrams in the doubled Hilbert space is also dual to each other. At $p>p_{c}^{(2)}$, both sides are characterized by non-vanishing correlation functions of operators, which corresponds to (left) the formation of mean-field for $Z_{v,u} Z_{v,l}$ operator or (right) condensation of a pair of $e$-anyons $e_u  e_l$. }
\end{figure}

\subsection{Lattice Model and Doubled Hilbert Space} \label{sec:choi}

\label{lattice}

We have shown that a decoherence channel whose Kraus operators are symmetric under $O(N)$ can drive an Ising-type phase transition that spontaneously breaks the $\mathbb{Z}_2^u \times \mathbb{Z}_2^l \subset O(N)_u \times O(N)_l$ symmetry down to the diagonal $\mathbb{Z}_2$ symmetry for quantities nonlinear in the density matrix. However, most physical quantities are linear in the density matrix, and this calls for a proper physical interpretation of these quantities. As we will see, our crossed-Ising transition is closely related to a better known information transition in the context of topological surface codes.  

In the following, we use another model to illustrate the essential physics of such a spontaneous symmetry breaking (SSB) from $\mathbb{Z}_2^u \times \mathbb{Z}_2^l$ to the diagonal $\mathbb{Z}_2$. Instead of starting with a state at quantum criticality, we consider a concrete lattice model with a trivially disordered state $| \Omega_0 \rangle = |+ \rangle^{\otimes N}$ on a $L \times L$ square lattice with qubits defined on vertices. We consider the following quantum channel as a symmetric local decoherence model:
\begin{align} \label{eq:noise_ising}
    \cE_{e=(v,v')}: \rho &\rightarrow (1-p)\rho + p Z_v Z_{v'} \rho Z_v Z_{v'},
\end{align}
where ``$e$" labels the edge between the nearest-neighbor pair of sites $v$ and $v'$. The decoherence channel is given as the composition of local channels, $\cE = \prod_e \cE_e$. To proceed further, we apply the Choi-Jamiolkowski isomorphism to map the density matrix into the pure state and decoherence channel into the operator in the doubled Hilbert space. Under this mapping, the Choi state of the pure state density matrix $\rho_0 = | \Omega_0 \rangle \langle \Omega_0 |$ is denoted as $\Vert \rho_0 \rAngle \equiv |\Omega_0\rangle |\Omega_0\rangle$, which is nothing but a vectorized density matrix, and the Choi operator of the decoherence channel is given as 
\begin{align} \label{eq:ising_dec}
    \hat{\cE} = \prod_{e=(v,v')} (1-2p)^{1/2} e^{\tau Z_{v,u} Z_{v',u} Z_{v,l} Z_{v',l}}
\end{align}
where $\tanh \tau = p/(1-p)$. As illustrated in \figref{fig:bilayer}, the isomorphism effectively maps the density matrix into the pure state in the bilayer system. 
Now, when the system is subject to the decoherence, we can show that $\Vert \rho^D \rAngle = \hat{\cE} \Vert \rho_0 \rAngle$ is the ground-state wavefunction of the following \emph{local} Hamiltonian in the doubled-Hilbert space~\cite{nishimorilee, sptdecohere}:
\begin{align} \label{eq:2d_doubleIsing}
    H_\textrm{tot} &= H_u + H_l + H_\textrm{int} \nonumber \\
    H_{u(l)} &= - \cosh 2\tau \sum_v X_{v,u(l)}  \nonumber \\
    H_\textrm{int} &=  \sum_v \cosh^4 2\tau \prod_{v' \in v}  (1 - Z_{v,u} Z_{v',u} Z_{v,l} Z_{v',l} \tanh 2\tau ),
\end{align}
where $v'\in v$ means that $v'$ is a nearest-neighbor site of $v$. At $p=\tau=0$, it reduces into two decoupled disordered states. At $p>0$, the decoherence gives rise to the (local) coupling between two layers that may induce a phase transition into an SSB phase that breaks the $\mathbb{Z}_2^u \times \mathbb{Z}_2^l$ into the diagonal $\mathbb{Z}_2$ symmetry.
The (unnormalized) groundstate can be written as
\begin{align} \label{eq:wavefunction}
    \hat{\cE} \Vert \rho_0 \rAngle =  (1-p)^{2N_v} \sum_\bl (\tanh \tau)^\abs{\bl} | \rd \bl \rangle \otimes | \rd \bl \rangle
\end{align}
where the summation is taken over all string configurations $\bl$ on the edges of the square lattice, $| \rd \bl \rangle \equiv \prod_{v \in \rd \bl} Z_v | \Omega_0 \rangle$, and $N_v = L^2$ is the number of vertices in the square lattice. We stress that $\hat{\cE} \Vert \rho_0 \rAngle$ is normalized in the sense that its corresponding density matrix $\rho^D$ (under the Choi-Jamiolkowski isomorphism) is normalized with $\tr\{ \rho^D \} = 1$ in the single Hilbert space while it is unnormalized as a state in the doubled Hilbert space.
Then, it is straightforward to show that the norm of the wavefunction $\hat{\cE} \Vert \rho_0 \rAngle$, i.e., the purity of the density matrix $\tr\{(\rho^D)^2 \}$ is equivalent to the partition function of the 2d Ising model at the temperature $\beta = 2 \tau$, as explicitly derived in \appref{app:2dIsing}. Accordingly, at $\beta=0.441$, which corresponds to $p^{(2)}_c= 0.178$, the wavefunction in the doubled Hilbert space undergoes the transition of 2d Ising universality. Here the superscript in $p$ implies that the transition happens in the quantity that involves the product of two density matrices.

The SSB transition between the $\mathbb{Z}_2^u \times \mathbb{Z}_2^l$-symmetric phase and the phase with only the diagonal $\mathbb{Z}_2$ symmetry in the doubled Hilbert space is captured by the susceptibility $\chi$ to an external symmetry-breaking field coupled to $O = \sum_{v} Z_{v,u} Z_{v,l}$ (in the doubled Hilbert space):
\begin{align} \label{eq:susceptibility}
    &\chi  \equiv \frac{1}{L^2} \Big( \lAngle O^2 \rAngle - \lAngle O \rAngle^2 \Big) \nonumber \\
    &\,= \frac{1}{L^2} \bigg( \frac{\lAngle \rho^D \Vert O^2  \Vert \rho^D \rAngle }{ \lAngle \rho^D \Vert  \rho^D \rAngle} -  \frac{\lAngle \rho^D \Vert O \Vert \rho^D \rAngle^2 }{ \lAngle \rho^D \Vert  \rho^D \rAngle^2} \bigg) \nonumber \\
    &\,=\sum_{v,v'} \frac{\tr\{ \rho^D Z_{v} Z_{v'} \rho^D Z_{v} Z_{v'} \}}{L^2\tr\{(\rho^D)^2 \} } - \left(\frac{ \sum_{v} 
 \tr\{ \rho^D Z_{v}  \rho^D Z_{v}  \}}{L\, \tr\{(\rho^D)^2 \} } \right)^{\hspace{-2pt}2} 
\end{align}
which is given by the summation over connected correlation functions of the order parameter $Z_{v,u} Z_{v,l}$ of this SSB transition in the doubled Hilbert space, related to the crossed-Ising correlation function $C^{(2)}_{XI}$ in \eqnref{eq:crossed_Ising}. $\chi$ is closely related to the crossed-Ising correlation functions in \eqnref{eq:crossed_Ising}. $\chi$ diverges as $|p-p_c^{(2)}|^{-7/4}$ at the SSB transition. The exponent $7/4$ is the signature of the 2d Ising universality class of this SSB transition.
Note that $\lAngle O \rAngle$ in the definition of $\chi$ is the order parameter that takes a non-zero value in the symmetry-broken phase and should be evaluated with $\delta=0^+$ after the thermodynamical limit is taken. We elaborate more on this in the next section.
 
{In the doubled Hilbert space, the definition of $\chi$ and the associated notion of the external symmetry-breaking field are standard for generic condensed-matter systems with global symmetries. In Sec. \ref{sec:CorrelFunc_PhysMeaning}, we explain their physical meaning at the level of the density matrix $\rho^D$.

\subsection{Correlation Functions and Physical Meanings}
\label{sec:CorrelFunc_PhysMeaning}
In the previous sections, we have defined several correlation functions or physical quantities which are nonlinear in decohered density matrices. Accordingly, these quantities are not directly accessible from experiments, raising questions about what they really mean physically. 
In order to connect these quantities to experiments, we consider the signatures of the transition in the probability distribution of measurement outcomes. 
More precisely, we will show that $\chi$ in \eqnref{eq:susceptibility} corresponds to the sensitivity of the decohered mixed state against small perturbations. This, in turn, is related to the amount of information that can be obtained from measuring the mixed state.

To proceed, we define the notion of distance between two different density matrices as the following:
\begin{align} \label{eq:div}
    D_\textrm{J}(\rho,\sigma) \equiv \tr( \rho \log \rho - \rho \log \sigma + \sigma \log \sigma - \sigma \log \rho  ).
\end{align}
This quantity is the quantum generalization of the Jeffreys divergence~\cite{Jeffreys1948} (or symmetrized quantum relative entropy), which quantifies the distance between two density matrices. It satisfies good properties to be a valid metric between mixed states: $(i)$ non-negative, $(ii)$ vanishing if and only if$\rho=\sigma$, and $(iii)$ monotonically decreasing under the application of quantum channels.

In order to discuss the ``sensitivity'' of the decohered density matrix, we define an infinitesimal variation of the original density matrix under symmetry breaking ``perturbation'', defined as the following quantum channel:
\begin{align} \label{eq:weak_meas}
    &{\cal M}_{\delta, v} : \rho \rightarrow (1-\delta) \rho + \delta Z_v \rho Z_v  \nonumber \\
    &\rho_\delta \equiv {\cal M}_\delta[\rho], \quad {\cal M}_\delta \equiv \prod_v {\cal M}_{\delta,v}.
\end{align}
which amounts to the application of weak measurement in the $Z$ basis. Hence, $\delta$ also serves as a symmetry-breaking field in the doubled Hilbert space. Then, $D(\delta) \equiv D_\textrm{J}(\rho^D, \rho^D_\delta)$ quantifies the difference between the unperturbed and perturbed decohered density matrices. Accordingly, how fast the distance changes with $\delta$ tells how sensitive the decohered system is against weak measurement in the $Z$ basis (or how informative the $Z$ measurement is). This is captured by the second derivative of the distance, which is related to the (classical) Fisher information as ${\cal F} \equiv \rd_\delta^2 D(\delta)\big|_{\delta=0}$~\cite{Fisher2021}.

However, the expression in \eqnref{eq:div} is very challenging to evaluate. 
To proceed, we generalize the Jeffreys distance in a way similar to Ref.~\onlinecite{BaoChoi2019}:
\begin{align}
    D^{(n)}(\rho,\sigma) &\equiv \frac{1}{n-1} \Big( \log \tr(\rho^n) + \log \tr(\sigma^n) \nonumber \\
    & \qquad - \log \tr(\rho \sigma^{n-1}) - \log \tr(\sigma \rho^{n-1}) \Big) 
\end{align}
This $n$-th Jeffreys distance is symmetric and non-negative~\footnote{However, it does not monotonically decrease under quantum channels anymore.}. Furthermore, in the limit $n \rightarrow 1^+$, $D^{(n)} \rightarrow  D_\textrm{J}$. At $n=2$, we find that the expression behaves like an overlap between Choi states of two density matrices in the doubled Hilbert space description:
\begin{align} \label{eq:distn2}
    D^{(2)}(\rho, \sigma) \equiv -\log( \frac{\lAngle \rho  \Vert \sigma \rAngle }{ \sqrt{ \lAngle \rho  \Vert \rho \rAngle \cdot \lAngle \sigma  \Vert \sigma \rAngle } } ),
\end{align}
With this definition, we evaluate the derivatives of the ``distance'' between $\rho^D$ and its perturbed version $\rho^D_\delta$ with respect to $\delta$ (See \appref{app:Fisher}).  This quantity exhibits interesting behaviors when $\delta \rightarrow 0$ and $L\rightarrow \infty$. The two different orders of limit lead to different results which have natural interpretations from a standard condensed-matter perspective and from a quantum information perspective respectively.

In the limit  where the thermodynamical limit $L \rightarrow \infty$ is taken first and $\delta \rightarrow 0^+$ afterward, the conventional choice of limit for a condensed-matter system experiencing SSB,  the second-order derivative $\rd^2_\delta D^{(2)}$ produces a quantity proportional to the susceptibility:
\begin{align} \label{eq:dist_suscept}
    \lim_{\delta\rightarrow0^+}\lim_{L\rightarrow\infty} \frac{1}{L^2} \rd^2_\delta D^{(2)} 
    %\big|_{\delta=0^+} 
    =  \chi,
\end{align}
which diverges at the SSB phase transition. 
Based on the nature of this SSB transition and the interpretation of $\delta$ as the symmetry-breaking field, we conclude that the decohered density matrix undergoes a qualitative change (with respect to this distance) as a function of $p$. Beyond critical $p_c^{(2)}$, the density matrix changes significantly under weak measurements in the $Z$ basis.

However, when we consider taking the limit $\delta\rightarrow 0$ before the thermodynamical limit, the second-order derivative produces a quantity that aligns with the information-theoretic intuition: 
\begin{align} \label{eq:dist_info}
    \lim_{L\rightarrow\infty} \lim_{\delta\rightarrow0} \frac{1}{L^4} \rd^2_\delta D^{(2)}  
    = \sum_{v,v'} \frac{\tr\{ \rho^D Z_{v} Z_{v'} \rho^D Z_{v} Z_{v'} \}}{L^4  \tr\{(\rho^D)^2 \} } = M,
\end{align}
where $M$ can be interpreted as the expectation value of the squared order parameter, which vanishes in the paramagnetic phase but acquires a finite value in the SSB phase. \cmj{Since this SSB transition belongs to the 2d Ising universality class, the standard phenomenology of the Ising model can be directly translated into the language of our study. In an Ising ferromagnet}, the magnetic order of the ground state of the system is very sensitive to an infinitesimal external Zeeman field. This well-known feature corresponds to the sensitivity of the density matrix to the decoherence in the $Z_v$ basis (as in \eqnref{eq:weak_meas}), in the phase where the $\mathbb{Z}_2^u \times \mathbb{Z}_2^l$ symmetry is spontaneously broken down to the diagonal $\mathbb{Z}_2$. This is quantified by the value of $\rd^2_\delta D^{(2)} \big|_{\delta=0}$ (where $\delta$ is set to 0 before $L$ grows large) scaling as $L^4$ in the SSB phase, which aligns with the Fisher information of the GHZ state  in the context of quantum metrology~\cite{Friis_2017}.

It is straightforward to argue that the second derivatives of the distance $D^{(n>2)}$ in both limits  would also exhibit similar behavior across a certain critical value (that depends on $n$), particularly because $\tr( (\rho^D)^n )$ is mathematically equivalent to the partition function of the coupled $n$-copies of Ising model which is expected to undergo a transition from a paramagnetic to a ferromagnetic phase potentially at a different critical strength $p_c^{(n)} \geq p_c^{(2)}$. Accordingly, we expect the behavior of $\rd^2_\delta D^{(n)}$ to extrapolate in the limit $n \rightarrow 1$. Therefore, there should be two different phases separated by a critical point at strength $p = p_c^{(1)}$. This is indeed the case, as we will see by using a dual description of the model.

\subsection{Duality, Intrinsic transition, and Decodability}

Under the Kramers-Wannier (KW) duality, the $\mathbb{Z}_2$ paramagnet under symmetric decoherence maps to the toric code under dephasing noise. The toric code under dephasing noise, in turn, is well known to exhibit an information transition at critical noise strength, beyond which the quantum information (logical qubits) stored in the toric code is not decodable~\cite{kitaevpreskill}. 
This hints at an intimate connection between the criticality discussed above and a well-known information transition.

First, we illustrate how the doubled Hilbert space formalism provides us some insights into the transition in the decohered toric code. Under the KW duality (See \appref{app:KW}), the Hamiltonian in \eqnref{eq:2d_doubleIsing} maps to the two copies of toric code coupled by local anyon tunneling terms, as schematically shown in \figref{fig:bilayer}. 
At critical strength of the tunneling, the anyon condenses and the topological order reduces into a single toric code.
The critical behavior is associated with a Higgs transition signified by the development of the anyon condensation amplitude $\sim \lAngle \rho^D \Vert \hat{\gamma}_u \hat{\gamma}_l \Vert \rho^D \rAngle$, where $\hat{\gamma}_{u/l} = \prod_{e \in \gamma^\perp} Z_{e,u/l}$ creates a pair of $e$-anyon at the end of the string $\gamma^\perp$. In turn, this implies that $\hrho^D$ and $\hat{\gamma} \hrho^D \hat{\gamma}$ have an appreciable overlap and they become less and less distinguishable for $p > p_{c}^{(2)}$ in the doubled Hilbert space.

In this coupled toric code in the doubled Hilbert space, the existence of the criticality can be directly seen by calculating the norm of the wave function, which is equivalent to the purity of the decohered toric code. The said quantity is given by the partition function of the 2d Ising model at $\beta = \tanh^{-1}(1-2p)^2$ as shown in \appref{app:purity_toric}, which has a transition at $\beta = \ln(1+\sqrt{2})/2$. As expected from the KW duality, this transition from an ordered to disordered phase upon increasing $p$ coincides with the transition of the original Choi state \eqnref{eq:wavefunction} from a disordered to an ordered phase, giving rise to the same critical point $p_c^{(2)}=0.178$.

As a next step, we calculate the von Neumann entropy of the decohered toric code state, a quantity highly nonlinear in the density matrix that enters the expression in \eqnref{eq:div}. Following the calculations in \appref{app:toric_ent}, we show that the von Neumann entropy of the decohered density matrix is proportional to the free energy of the random bond Ising model along the Nishimori line~\cite{Nishimori, Nishimori2} at $\beta = \tanh^{-1}(1-2p)$, which is known to be critical at $p=0.1094$~\cite{Doussal1988}.
Therefore, the transition behavior indeed extends down to the $n\rightarrow 1$ limit  for a toric code under dephasing noise, and by the KW duality, for the original paramagnetic spin model under symmetric decoherence. As remarked, the transition behavior in the limit $n\rightarrow 1$ corresponds to the singularity of the Fisher information, a well-known information-theoretic quantity. 
We remark that this series of transitions for different $n$ is \emph{intrinsic} to the decohered density matrix (and the distance in use), as these transitions are associated with spontaneous symmetry breaking. Interestingly, the $n \rightarrow 1$ limit of this intrinsic transition coincides with the decodability transition point demonstrated in Ref.~\onlinecite{kitaevpreskill}, which is based on a certain decoding procedure. Such a connection may imply that the intrinsic transition point provides an upper bound for some information retrieval protocols to be successful in this setting.

\section{Non-local operators}

\label{nonlocal}

As we mentioned previously, without any postselection, the correlation function for local operators linear with the decohered density matrix should have the same scaling as the undecohered correlation function. But in this section, we will show that {\it nonlocal quantities} can still have qualitatively different behaviors even if we only consider expectation values linear with the density matrix.

\subsection{\texorpdfstring{$1d$}{1d} Quantum Rotor}

To illustrate the behavior of nonlocal operators under decoherence, let us start with the $1d$ systems with a description in terms of the quantum rotor, such as the spin-1/2 chain. In terms of the Abelian bosonization, the N\'{e}el and valence bond solid (VBS) order parameters of a spin-1/2 chain are represented as \beqn \left( N^x, \ N^y, \ N^z, \ V \right) \sim \left( \sin \theta, \ \cos \theta, \ \sin \phi, \ \cos \phi \right). \eeqn Under decoherence or weak measurement of (for example) local operator $e^{\ii \hphi}$, the density matrix becomes  \beqn && \hrho^D = \cE[\hrho_0], \ \ \ \cE = \prod_x \cE_x, \cr \cr \cE_x[\hrho_0] &\sim& (1 - p) \hrho_0 + \frac{p}{2} \ e^{\ii \hphi(x)} \hrho_0 e^{- \ii \hphi(x)} \cr\cr &+& \frac{p}{2} \ e^{- \ii \hphi(x)} \hrho_0 e^{\ii \hphi(x)}. \label{hrhoD} \eeqn $\hrho_0$ is the undecohered density matrix of the spin-1/2 chain, and $\hrho^D = \cE[\hrho_0]$ still keeps $\tr[\hrho^D] = 1$.

We can first evaluate the correlation function of local operator $O(r)$: \begin{widetext} \beqn \label{eq:channeldec}
&& C^D(r) = \tr \{ \hat{\rho}^D \hO(r) \hO(0) \} = \tr \{ \hrho_0 \cE [\hO(r) \hO(0)] \}, \cr\cr && \cE_x [\hO(r) \hO(0) ] \sim (1 - p) \hO(r) \hO(0) + \frac{p}{2} \ e^{\ii \hphi(x)} \hO(r) \hO(0) e^{- \ii \hphi(x)} + \frac{p}{2} \ e^{- \ii \hphi(x)} \hO(r) \hO(0) e^{\ii\hphi(x)} \eeqn \end{widetext} 
% For local bosonic order parameters, such as the N\'{e}el and VBS, although they may have nontrivial commutation with $e^{\ii \hphi(x) }$ at the vicinity of $x$, at long distance apart, $\hO(r)$ and $e^{\ii \hphi(x)}$ should commute. 
For local bosonic order parameters at position $r$, such as the N\'{e}el and VBS,  their nontrivial commutation with $e^{\ii \hphi(x) }$ would be exponentially suppressed by the separation $|r - x|$ at long distance, and thus for $|r - x| \gg 1$, $\hO(r)$ and $e^{\ii \hphi(x)}$ should commute.  
When $\hat{O} \sim \cos(\phi)$ or $\sin(\phi)$, since $e^{i\hat{\phi}(x)}$ always commute with $\hat{O}(0)$ and $\hat{O}(r)$, $\cE_x[\hat{O}(r)\hat{O}(0)] = \hat{O}(r)\hat{O}(0)$, and the decoherence does not affect the correlator. When $\hat{O} = N^x$ or $N^y$, the decoherence channel does not commute with $\hat{O}(r)$ only in the vicinity of $r$. However, ($i$) the field $(N^x, N^y)$ is already the primary field with the lowest scaling dimension that forms an irreducible representation of the $\U(1)$ rotation symmetry generated by $\sum_{x} S^z_{x}$ of the spin chain ($\theta \mapsto \theta + \alpha$), and ($ii$) the decoherence channel explicitly preserves this $\U(1)$ symmetry, $e^{\ii \hat{\phi}(r)} (N^x_r, N^y_r) e^{- \ii \hat{\phi}(r)} $ belong to the same representation as $(N^x_r, N^y_r)$ and would not generate any field with lower scaling dimension. Therefore, under decoherence the correlation functions of local operators $(N^x, N^y, N^z, V)$ acquire only a constant amount of multiplicative factor aside from potential subleading power-law contributions, implying that they should still have the same power-law with scaling dimensions of the undecohered spin-1/2 chain
% When $\hat{O} = N^x$ or $N^y$, Although the decoherence channel do not commute with $\hat{O}(r)$ in the vicinity of $r$, because ($i.$) $(N^x, N^y)$ is already the primary field with the lowest scaling dimension that forms a irreducible representation of the $\U(1)$ rotation symmetry generated by $\sum_{x} S^z_{x}$ of the spin chain, and ($ii.$) the decoherence channel explicitly preserves this $\U(1)$ symmetry, $e^{\ii \hat{\phi}(r)} (N^x_r, N^y_r) e^{- \ii \hat{\phi}(r)} $ belong to the same representation as $(N^x_r, N^y_r)$ and won't generate any field with lower scaling dimension. Therefore, the correlation functions of local operators $(N^x, N^y, N^z, V)$ acquire only a constant amount of multiplicative factor under decoherence, implying that they should still have the same power-law with scaling dimensions of the undecohered spin-1/2 chain.
\footnote{The conclusion here should generally hold for decoherence channel and correlation functions of local operators, i.e. as long as ({\it i}) the local operator is the primary field with the lowest scaling dimension of a certain representation of the symmetry, and ({\it ii}) the decoherence channel explicitly preserves this symmetry (with a ``doubled'' symmetry condition) and commute with the local operator except for the close vicinity of the operator, then the decoherence should not change the long distance scaling of the correlation function. }.

However, for nonlocal operators, the situation can be very different. A particular family of nonlocal operators is the {\it disorder operators}~\cite{disorder,cat,odo1,meng1,odo2,meng2,meng3,meng4}~\footnote{It was also called by other names, such as patch operator~\cite{cat}, and order diagnosis operator~\cite{odo1,odo2}, etc.}, which have attracted great interests recently. These operators have been used as an auxiliary diagnosis for the states of matter, especially for critical states of matter. For a $1d$ quantum rotor, if we view $\hphi$ as the phase angle of a local boson creation operator, then $(\nabla_x \htheta)/2\pi$ is the boson density $\hn_{\phi}$. The following operator is called a disorder operator \beqn \tilde{O}_r &=& \exp\left( \ii \alpha \int_0^r dx \ \hn_\phi(x)\right) \cr\cr &=& \exp\left( \ii \frac{\alpha}{2\pi} (\htheta(r) - \htheta(0)) \right). \eeqn  If $\hphi$ carries a full U(1) symmetry, then $\alpha \in \mathbb{R}$; if $\hphi$ only carries a $\mathbb{Z}_N$ symmetry, then $\alpha = \frac{2\pi k}{N}$ with $k \in \{1,...,N\}$, as $\hn_\phi$ is defined modulo $N$.

For example, for the spin-1/2 chain, if we view $\hphi$ as a local operator, then $e^{\ii \htheta/2}$ is a disorder operator of $\hphi$, and $e^{\ii \htheta/2}$ plays two roles simultaneously: it first creates a fractionalized spin-1/2 excitation (i.e. a spinon), it also creates a domain wall of the VBS order parameter, i.e. $e^{\ii \htheta(r)/2}$ shifts $\hphi(x) \rightarrow \hphi(x) +\pi$ for $x < r$. Indeed, it is well-known that a spin-1/2 is localized at the domain wall between two VBS orders. We can evaluate the correlation function of $e^{\ii \htheta/2}$ for the decohered density matrix:  \beqn C^D(r) &=& \tr \{ \hat{\rho}^D e^{\ii \htheta(r)/2} \ e^{- \ii \htheta(0)/2} \} \cr\cr &\sim& e^{ - r/\xi} \tr \{ \hrho_0 \ e^{\ii \htheta(r)/2} \ e^{- \ii \htheta(0)/2 } \} , \eeqn where $\tr \{ \hrho_0 e^{\ii \htheta(r)/2} \ e^{- \ii \htheta(0)/2} \}$ is the correlation function of $e^{\ii \htheta/2}$ for undecohered spin-1/2 chain, and the ``correlation length" $\xi$ is $ \xi \sim - 1/\ln (1 - 2p) $ for small $p$. 

Hence our calculation for the $1d$ quantum rotor system implies that, although the disorder operator has a power-law correlation with the absence of decoherence, it can be rendered short-ranged under decoherence. As we will show in the next subsection, similar behavior of the disorder operator happens in higher dimensions as well. The spin-1/2 chain is also an example of spin liquid with fractionalized spinon excitations since the spinon correlation function decays as a power-law in the undecohered spin-1/2 chain. But under decoherence or weak measurement on the VBS order parameter the spin chain {\it loses its fractionalization}, as the spinon operator decays exponentially. This can be intuitively understood as the fact that, if the VBS operator is ``measured", in each measurement outcome the system is pinned to a certain particular VBS pattern, which leads to confinement in this measurement outcome. The confinement persists even if we average over all measurement outcomes.

\subsection{\texorpdfstring{$(2+1)d$}{(2+1)d} quantum critical points with a U(1) or \texorpdfstring{$\mathbb{Z}_N$}{ZN} symmetry}

Now let us consider a $(2+1)d$ QCP or CFT with a global U(1) symmetry, or $\mathbb{Z}_N$ symmetry that can be embedded into a U(1) that emerges in the infrared. This U(1) symmetry is dual to a noncompact U(1) gauge field~\cite{peskindual,halperindual,leedual}. We always turn on decoherence on the scalar boson creation operator which carries the U(1) charge, or equivalently the monopole operator of the dual U(1) gauge field. The decohered density matrix takes the same form as Eq.~\ref{hrhoD}: \beqn && \hrho^D = \cE[\hrho_0], \ \ \ \cE = \prod_\vect{x} \cE_\vect{x}, \cr \cr \cE_\vect{x} [\hrho_0] &\sim& (1 - p) \hrho_0 + \frac{p}{2} \ e^{\ii \hphi(\vect{x})} \hrho_0 e^{- \ii \hphi(\vect{x})} \cr\cr &+& \frac{p}{2} \ e^{- \ii \hphi(\vect{x})} \hrho_0 e^{\ii \hphi(\vect{x})}. \eeqn Here $e^{\ii \hphi(\vect{x})}$ is the monopole operator, which creates a scalar boson, or a $2\pi$ gauge flux at location $\vect{x}$.

We evaluate the expectation value of the following quantity defined for a closed loop $\cC = \rd \cA$:  \beqn \tilde{O}_\cC =  \exp\bigg( \sum_{\vect{x} \in \cA, \partial \cA = \cC} \frac{\ii 2\pi}{N} \hn_{\phi}(\vect{x})  \bigg). \eeqn In the undecohered density matrix, and in the dual formalism, this quantity reduces to the evaluation of the Wilson loop: \beqn \langle \tilde{O}_\cC \rangle &\sim& \langle \exp\bigg( \frac{\ii}{N} \oint_\cC d\vect{x} \cdot \hat{\vect{a}}(\vect{x}) \bigg) \rangle, \eeqn and as was shown previously, it should obey a perimeter law, with a universal logarithmic contribution from the sharp corners of the loop $\cC$~\cite{odo2,meng2,williamcorner}. The coefficient of the universal logarithmic contribution arising from the corner is proportional to the universal conductivity of the scalar boson current at the $(2+1)d$ CFT in the AC limit $\omega/T \rightarrow \infty$.

However, under decoherence, the operator $\tilde{O}_\cC$ will shift $\hphi(\vect{x})$ by angle $2\pi/N$ for $\vect{x} \in \cA$. Hence we expect the decoherence to change the behavior of $\langle \tilde{O}_\cC \rangle$ significantly: \beqn \langle \tilde{O}_\cC \rangle \sim \left( (1-p) + p \cos\left(\frac{2\pi}{N}\right) \right)^\cA, \eeqn namely $\langle \tilde{O}_\cC \rangle$ should now decay with an area law. Just like the $1d$ example discussed in the previous subsection, an area law decay of the Wilson loop is a sign of {\it confinement}. It means that the vortex of the $\U(1)$ boson, which is also the gauge charge of the dual gauge field $\hat{\vect{a}}$, should be confined under decoherence of the scalar boson creation operator. Here we would like to remark that, one of the tools for diagnosing fractionalization and deconfinement is the dynamic structure factor, where the fractionalization would lead to a continuum~\cite{dynamic1d,dynamic2d1,dynamic2d2}. Computing real-time dynamics is beyond the current set-up of our current manuscript as it requires the formalism that involves the Lindbladian. Here we use the behavior of the Wilson loop as the sign of confinement/deconfinement.

\section{summary and future directions}

In this study, we examined the effects of decoherence and weak measurement on quantum critical points in $(2+1)d$ space-time. We found that this problem is mathematically equivalent to the boundary or defect criticality of $(2+1)d$ conformal field theories, which have been extensively researched in recent years. Our results indicate that when a QCP is exposed to decoherence or weak measurement, observers may observe peculiar behaviors, including the extraordinary-log correlation recently discovered in the context of boundary criticality. Additionally, as the strength of decoherence or weak measurement increases, the system can experience an information-theoretic transition that is captured by quantities nonlinear in the decohered density matrix. This transition is linked to spontaneous symmetry breaking when we consider quantities to the $n$-th power of the density matrix; in particular using the ``doubled formalism", we provided a specific example where the transition belongs to the $2d$ Ising universality class for $n=2$. By KW duality, this transition is connected to the \emph{error threshold} transition in the decohered toric code, which is another type of information-theoretic transition in quantum systems. 
% In this work, we studied quantum critical points (especially in $(2+1)d$ space-time) exposed to decoherence or weak measurement. It turns out that this problem is mathematically mapped to the boundary or defect criticality of $(2+1)d$ conformal field theories, which has been the subject of significant interest and research efforts in recent years. We find that for a QCP under decoherence/weak measurement, an observer is expected to observe peculiar behaviors including the extraordinary-log correlation discovered recently in the context of boundary criticality. We also find that as the strength of decoherence/weak measurement increases, the system can undergo an information-theoretic transition that is captured by quantities that are nonlinear with the decohered density matrix. This information transition is associated with a spontaneous symmetry breaking when we consider quantities to the $n-$th power of the density matrix; in particular, using the ``doubled formalism", we show that for $n = 2$ this transition belongs to the $2d$ Ising universality class.

There are many related directions that are very much worth exploring in the future. We list two such directions as follows:

{\it (1) Unconventional quantum criticality under decoherence:} 
It is known that in the world of quantum many-body systems, there are two types of quantum critical points: conventional and unconventional. Conventional QCPs correspond to quantum phase transitions between a disordered state that can be adiabatically connected to a direct product state, and a state that spontaneously breaks certain symmetry. This type of QCPs has classical analogs such as the Wilson-Fisher fixed points discussed in this work. Well-understood examples of such conventional QCP can be found in Ref.~\cite{sachdevbook}, including the QCP of the transverse field quantum Ising model, which is between the Ising ordered phase that spontaneously breaks the Ising symmetry, and a symmetric disordered phase that is adiabatically connected to a direct product state. For a two-dimensional quantum Ising model, the QCP is described by the $3D$ Ising Wilson-Fisher fixed point. Another well-known example of such conventional QCP is the quantum phase transition between the superfluid phase and the Mott insulator phase in the Bose Hubbard model at integer filling~\cite{bosehubbard}, which belongs to the 3D XY Wilson-Fisher universality class if the model is built on a two-dimensional lattice.  
%But it is known that there is another large class of {\it unconventional} QCPs without a simple classical analogue~\cite{uqcp}. 

On the other hand, unconventional QCPs are those that do not have a simple classical analog and may involve transitions between two ordered phases with different symmetries, or between an ordered phase and a topological order. The most well-known example of unconventional QCPs is the deconfined QCP~\cite{deconfine1,deconfine2}, which has many desirable phenomena such as deconfinement and a duality web as was summarized in Ref.~\onlinecite{SO5}.  There is another large class of unconventional QCPs which involve quantum disordered states that cannot be adiabatically connected to a direct product state, such as the topological orders. One early example of such quantum phase transitions can be found in Ref.~\cite{wenwu}.  It is reasonable to expect that the unconventional QCPs under decoherence can also be mapped to certain boundary criticality problems, and it is going to be an unusual boundary criticality with unconventional QCP in the bulk. As we have already seen in the current work, decoherence may be at odds with deconfinement, as deconfinement is often signified by nonlocal operators such as the disordered operators or the Wilson loops, whose behavior can be strongly affected by decoherence. It would be interesting to study the fate of unconventional QCPs under decoherence in general in the future.

{\it (2) The Strange Correlator:}
The notion of a strange correlator was originally proposed as a tool to diagnose SPT states using their bulk wave functions~\cite{YouXu2013}, rather than edge states. The strange correlator is defined as the following quantity \beqn C^S(\vect{r}) = \frac{\langle \Omega | \hat{O}(\vect{0}) \hat{O}(\vect{r}) | \Psi \rangle}{\langle \Omega | \Psi \rangle}, \label{strange}\eeqn where $|\Psi\rangle$ is the wave function that awaits diagnosis, and $|\Omega\rangle$ is the trivial direct product disordered state with the same symmetry $G$ and Hilbert space as $|\Psi\rangle$. $\hat{O}$ is an order parameter that carries a nontrivial representation of $G$. The arguments given in Ref.~\onlinecite{YouXu2013} suggest that, although the ordinary correlation functions in both $|\Psi\rangle$ and $|\Omega\rangle$ must be short-ranged, this strange correlator Eq.~\ref{strange} must have either long-ranged or power-law correlation, for $1d$ and $2d$ states. In the past decade, the strange correlator has been used as a tool for both conceptual understanding and numerical diagnosis for SPT states and also topological states~\cite{scwierschem1,scwierschem2,scwierschem3,sczohar1,scmeng1,sczohar2,scmeng2,scwei,scmeng3,scwierschem4,sczhong,scscaffidi,scfrank1,scfrank2,scfrank3,scfrank4,scfrank5,schsieh,scfan,scsagar,scmeng4}.

One of the future directions worth pursuing is the strange correlator between a quantum critical state $|\Omega\rangle$, and an SPT wave function $|\Psi\rangle$. Let us still focus on two-dimensional systems. Using the formalism developed in this work, this problem may be mapped to the $2d$ interface between a quantum critical point on the temporal domain $\tau < 0$, and an SPT state on the other domain $\tau > 0$ in the Euclidean space-time path-integral. Under space-time rotation, the strange correlator is mapped to the spatial interface between a quantum criticality and an SPT state. This kind of interface has two types of boundary effects: the boundary states arising from the bulk topology, and also the boundary criticality originating from the bulk critical modes. This is a subject under very active research lately, both theoretically and numerically~\cite{groveredge,zhang1,zhang2,stefan1,stefan2,edgexu2,shang1}. In particular, some novel interface criticality especially a $(1+1)d$ deconfined quantum critical point was identified in the literature~\cite{edgexu2}. One potentially highly interesting direction in the future is to analyze the strange correlator (and its generalized form defined in Ref.~\onlinecite{sptdecohere}) between quantum criticality and SPT state, and explore the possible novel phenomena, especially when either the bulk quantum critical state $|\Omega\rangle$, or the SPT state $|\Psi\rangle$, or both are under decoherence.

\acknowledgements 

We thank Ehud Altman, Soonwon Choi, Matthew P. A. Fisher, Sam Garrett, Yi-Zhuang You for inspiring discussions and previous collaborations.
J.Y.L. is supported by the Gordon and Betty Moore Foundation under the grant GBMF8690 and by the National Science Foundation under the grant PHY-1748958.
C. X. acknowledges the support from the Simons Foundation
through the Simons Investigator program. C.-M. J. is supported by a faculty startup grant at Cornell University.

\emph{Note Added:} While finishing up this work, we became aware of an independent related work~\cite{Bao2023,Fan2023}, which should appear on arXiv on the same day as our work.

\appendix

\section{Toric code under decoherence}

The goal of this section is to show that the entanglement entropy of the toric code state under dephasing noise is given by the free energy of the random bond Ising model along the Nishimori line. 
We will show that such a decohered state is dual to the disordered product state under $\mathbb{Z}_2$ symmetric decoherence channel in \eqnref{eq:ising_dec}.

First, the toric code Hamiltonian is defined as
\begin{align} \label{eq:toric}
        H &= - \sum_{v} \prod_{e \ni v } {Z}_e - \sum_{p} \prod_{e \in p} X_e \nonumber \\
        &= - \sum_v A_v - \sum_p B_p
\end{align}
The ground state is characterized by $A_v = B_p = 1$. Furthermore, on  the torus, the ground state is 4-fold degenerate with two logical qubits. Logical qubits reside on the space where the following effective Pauli operators act on $\bm{C}^X_i \equiv \prod_{e \in C_i} X_e$ and $\bm{C}^Z_i \equiv \prod_{e \in C^\perp_i} Z_e$ where $C_i$ is a cycle along the $i$-th axis; while $C_i$ is along the bond, $C_i^\perp$ crosses the bond. Note that $\{\bm{C}^X_1, \bm{C}^Z_2 \} = \{\bm{C}^Z_1, \bm{C}^X_2 \} = 0$, while $[\bm{C}^Z_i, \bm{C}^X_i] = 0$. As an example we choose to study one of the four ground states denoted as $|\psi_\tc \rangle$, whose pure state density matrix is $\rho_\tc = |\psi_\tc \rangle \langle \psi_\tc |$. The ground state is the eigenstate of the qubits $\bm{C}^X_i |\psi_\tc \rangle = a_i | \psi_\tc \rangle$, with $a_1 = a_2 =1$.
%be the ground state of the above Hamiltonian, and let $\rho_\tc = |\psi_\tc \rangle \langle \psi_\tc |$. Here, let $\bm{C}^X_i |\psi_\tc \rangle = a_i | \psi_\tc \rangle$. We choose $a_1 = a_2 =1 $.

\subsection{Decomposition} \label{app:toric_dec}
 
We consider the toric code ground state decohered under the following channel:
\begin{align} \label{eq:dephasing}
    \cE_e: \rho &\rightarrow (1-p)\rho + p Z_e \rho Z_e, \quad \cE = \prod_e \cE_e
\end{align}
In order to understand the structure of $\cE[\rho_\tc]$, first we evaluate the matrix elements of the decohered toric code density matrix. To do the job, consider $\rho_{\bs,\bs'} \equiv | \Omega_{\bs'} \rangle \langle \Omega_\bs | $ where $| \Omega_{\bs} \rangle$ is a generic product state characterized by $\bs = \{s_e \}$, $s_e =\pm 1$:
\begin{align}
    | \Omega_\bs \rangle \equiv \prod_{e} Z^{(1-s_e)/2} |+\rangle^{\otimes {2N_v}},
\end{align}
where $N_v = L^2$ is the number of vertices. Then 
\begin{align} \label{eq:element}
    \langle \Omega_\bs | \cE[\rho_\tc] | \Omega_{\bs'} \rangle &= \tr( \rho_{\bs,\bs'} \cE[\rho_\tc]) =  \tr( \cE[ \rho_{\bs,\bs'} ] \rho_\tc)
\end{align}
where $\cE[ \rho_{\bs,\bs'} ]$ is given as
\begin{align}
     \rho_{\bs,\bs'}   &=  \frac{1}{2^{{2N_v}}}   \prod_{e} (1 + s_e X_e) Z_e^{(1-s_e s'_e)/2} \nonumber \\
     \cE[ \rho_{\bs,\bs'} ] &=  \frac{1}{2^{{2N_v}}}   \prod_{e} (1 + s_e (1-2p) X_e) Z_e^{(1-s_e s'_e)/2} 
\end{align}
For $ \tr( \cE[ \rho_{\bs,\bs'} ] \rho_\tc)$  not to vanish, $\rd(\bs \cdot \bs')= 0$ so that product of $Z_e$ does not create anyons. In such a case, the product of $Z$-strings always commutes with a loop of $X$-strings along the bond. 
In fact, we can show that
\begin{align}
 \langle \psi_\tc | \prod_{e \in l} X_e \prod_e Z_e^{(1-s_e s_e')/2} | \psi_\tc \rangle = F(l, \bs \cdot \bs')  \delta_{\rd l,0} \delta_{\rd(\bs\cdot\bs'),0}   
\end{align}
where $\bs\cdot \bs'$ defines a non-trivial link configuration along the \emph{dual link} whenever it takes a negative value.  
For $\langle \bm{C}_{1,2}^{X,Y,Z} \rangle  = c^{x,y,z}_{1,2}$ we have
\begin{align}
    F(l, \bs \cdot \bs') = \langle \psi_\tc | \bm{C}_{h(l)}^X \bm{C}_{h(\bs \cdot \bs')}^Z | \psi_\tc \rangle 
\end{align}
where $h(l) \in \pi_1(\mathbb{T}^2)$ is the element of the homotopy group of the torus. For the simplest case where $\bm{C}_{1,2}^X = 1$, we remark that $F(l, \bs \cdot \bs')$ is non-zero if and only if$h(\bs,\bs')$ is trivial.
Then, the overlap in \eqnref{eq:element} is given as
\begin{align}
    \tr( \cE[ \rho_{\bs,\bs'} ] \rho_\tc) & = \frac{1}{2^{{2N_v}}} \sum_l (1-2p)^{\abs{l}}  \prod_{e\in l} s_e  \nonumber \\
    &\times  \langle \psi_\tc | \prod_{e \in l} X_e \prod_e Z_e^{(1-s_e s_e')/2} | \psi_\tc \rangle \nonumber \\
    &= \frac{\delta_{h(\bs \cdot \bs'), \bm{1}} }{2^{N_v} (2 \cosh \beta)^{{2N_v}}} Z_{\textrm{RBIM}}[\bs,\beta]
\end{align}
where the summation is over all possible edge configuration $l$, $\tanh \beta = (1-2p)$, and
\begin{align}
    Z_{\textrm{RBIM}}[\bs,\beta] \equiv \sum_{\bsigma} \prod_{e=(v,v')} e^{\beta s_e \sigma_{v} \sigma_{v'}},
    % \nonumber \\
    % e^{\beta m} =  \cosh \beta (1 + m \tanh \beta) \quad \forall m \in \{1,-1\} 
\end{align}
which turns out to be the partition function of an Ising model with random signs, specified by $\{s_e\}$, in the near-neighbor spin-spin interaction.
At $p=0$, $\beta^{-1} = 0$, i.e., zero temperature limit, It has an interesting consequence: the matrix element $\langle \Omega_\bs | \cE[\rho_\tc] | \Omega_{\bs'} \rangle$ vanishes unless $\bs$ and $\bs'$ belong to the same equivalence class, i.e., $\rd(\bs \cdot \bs') = 0$. Furthermore, if $\bs \sim \bs'$, then $Z_{\textrm{RBIM}}[\bs,\beta] = Z_{\textrm{RBIM}}[\bs',\beta]$. Accordingly, $\cE[\rho_\bs]$ is block-diagonal. Therefore, $\cE[\rho_\tc]$ decomposes as the following:
\begin{align}
    \cE[\rho_\tc] &=   \sum_{\bs,\bs'} \rho_{\bs,\bs'} \tr(\rho_{\bs,\bs'} \cE[\rho_\tc]) \nonumber \\
    &=  \frac{1}{2^{N_v} (2 \cosh \beta)^{{2N_v}}} \sum_\bmm Z_{\textrm{RBIM}}[\bs_\bmm,\beta] \, \rho_{\bmm}.
\end{align}
where the summation is taken over the equivalence class of $\bs$, denoted by $\bmm$; the equivalence class is defined as $\rd(\bs \cdot \bs') = 0$. $\bs_\bmm$ is the representative of the equivalence class $\bmm$, and $\rho_\bmm$ is defined as 
\begin{align}
\rho_\bmm \equiv \sum_{\bs,\bs' \sim \bs_\bmm} |\Omega_\bs \rangle \langle \Omega_{\bs'}|.    
\end{align}
Therefore, the decohered density matrix has the following block-diagonal structure (in $X$ basis)
\begin{align}
    \cE[\rho_\tc] =   \setlength{\arraycolsep}{0pt}
  \begin{bmatrix}
    \,\fbox{$\bm{B}_{1}$} & 0 & 0 & \,\,\cdots\,  \\
    0 & \fbox{$\bm{B}_{2}$} & 0 & \,\,\cdots \\
    0 & 0 & \fbox{$\bm{B}_{3}$} & \,\,\cdots \\
    \vdots & \vdots & \vdots  & \,\,\ddots
  \end{bmatrix}
\end{align}
where each block is $2^{N_v-1}$ by $2^{N_v-1}$ dimensional matrix labeled by the equivalence class $\bmm$ and its entries are all equal, i.e., 
\begin{align}
    \bm{B}_i \propto  
  \setlength{\arraycolsep}{2pt}
  \underbrace{ \begin{bmatrix}
    1 & 1 & 1&\,\cdots  \\
    1 & 1 & 1 &\,\cdots \\
    1 & 1 & 1 &\,\cdots \\
    \svdots & \svdots & \svdots &\,\sddots
  \end{bmatrix} }_{\displaystyle 2^{N_v-1}}
  \left.\vphantom{
  \begin{bmatrix}
    1 & 1 & 1&\,\cdots  \\
    1 & 1 & 1 &\,\cdots \\
    1 & 1 & 1 &\,\cdots \\
    \svdots & \svdots & \svdots &\,\sddots
  \end{bmatrix}}
  \right\} 2^{N_v-1} = 2^{N_v-1} |\phi_\bmm \rangle \langle \phi_\bmm |
\end{align}
where $|\phi_\bmm\rangle = \frac{1}{\sqrt{2^{N_v-1}}} \sum_{\bs \sim \bs_\bmm} |\Omega_\bs \rangle$. There are total $2^{N_v+1}$ equivalence classes originated from $(N_v-1)$ independent stabilizers $B_p$ and two logical operators $\bm{C}^X_{i}$.

\subsection{Entanglement Entropy} \label{app:toric_ent}

After reorganizing terms, we get
\begin{align} \label{eq:tc_decomposition}
    \rho_\tc^D &= \sum_\bmm p_\bmm |\phi_\bmm \rangle \langle \phi_\bmm |, \quad p_\bmm = \frac{Z_\textrm{RBIM}[\bs_\bmm,\beta]}{2 \cdot (2\cosh \beta)^{{2N_v}}} .
\end{align}
Note that the entanglement entropy has a very interesting structure:
\begin{align}
    S &= -\tr( \rho_\tc^D \ln \rho_\tc^D ) \nonumber \\
    &= - \sum_\bmm p_\bmm \log p_\bmm \nonumber \\
    &\propto - \sum_\bmm Z [\bs_\bmm,\beta] \log Z [\bs_\bmm,\beta]
\end{align}
which is nothing but a disorder averaged random bond Ising model's free energy along the Nishimori line, whose transition point is located at $p_{c} = 0.1094$~\cite{Doussal1988}. Therefore, there is an \emph{intrinsic} phase transition of the entanglement entropy of the decohered toric code state at $p_{c} = 0.1094$, which coincides with the decodability transition point obtained in Ref.~\onlinecite{kitaevpreskill}.

\subsection{Purity} \label{app:purity_toric}

Interestingly, the purity of the decohered density matrix maps to the partition function of the Ising model: (Here $C \equiv ({2^2 \cdot (2 \cosh \beta)^{4N_v}})^{-1}$):
\begin{align}
        &\tr( (\rho_\tc^D)^2 ) =  C \sum_{\bmm} Z^2_\textrm{RBIM}[\bs_\bmm,\beta]  = C \sum_\bs \frac{Z^2_\textrm{RBIM}[\bs,\beta]}{2^{N_v-1}}  \nonumber \\
    &= \frac{C (\cosh \beta)^{4N_v} }{2^{N_v-1}} \sum_\bs \sum_{\bsigma} \sum_{\bsigma'} \sum_{L_1,L_2} (1-2p)^{|L_1|+|L_2|} \nonumber \\
    &\qquad \times \prod_{e \in L_1} s_e \prod_{e' \in L_2} s_{e'} \prod_{i \in \rd L_1} \sigma_i \prod_{j \in \rd L_2} \sigma'_j
\end{align}
where $L_i$ is the link configuration defined on the square lattice. This expression can be further simplified by the following:
\begin{align} \label{eq:transition}
    \tr( (\rho_\tc^D)^2 ) &= \frac{1}{2^{4N_v + N_v +1}}  \sum_{\bsigma} \sum_{\bsigma'} \sum_{L_1,L_2} (1-2p)^{|L_1|+|L_2|} \nonumber \\
    &\qquad \times 2^{{2N_v}} \delta_{L_1, L_2} \prod_{i \in \rd L_1} \sigma_i \prod_{j \in \rd L_2} \sigma'_j  \nonumber \\
    &= \frac{1}{2^{4N_v + N_v +1}}  \sum_{L_1,L_2} (1-2p)^{|L_1|+|L_2|} \cdot 2^{{2N_v}} \delta_{L_1, L_2} \nonumber \\
    &\qquad \times 2^{N_v} \delta_{\rd L_1, 0} \cdot 2^{N_v} \delta_{\rd L_2, 0} \nonumber \\
    &= \frac{1}{2^{N_v+1}} \sum_{L} (1-2p)^{2\abs{L}} \delta_{\rd L, 0} \nonumber \\
    &= \frac{Z_\textrm{FIM}[\beta']}{2  (2 \cosh \beta')^{{2N_v}} }, \quad \tanh \beta' = (1-2p)^2
\end{align}
where $Z_\textrm{FIM}[\beta'] = Z_\textrm{RBIM}[\bm{1},\beta']$ is the partition function of the ferromagnetic Ising model. In the last equality, we used that 
\begin{align}
    \sum_{\rd \gamma=0} (\tanh \beta)^{|\gamma|} \prod_{e \in \gamma} s_e  = \frac{Z_{\textrm{RBIM}}[s,\beta]}{2^{N_v} (\cosh \beta)^{{2N_v}}}.
\end{align}
Since the ferromagnetic 2d Ising model has a transition at $\beta = \ln(1+\sqrt{2})/2 = 0.441$, correlation functions in the doubled Hilbert space (in the next section) would exhibit a critical behavior at $p^{(2)}_{c} = 0.178$. 
%$p_{c,(2)} = 0.178$.

% \subsection{$n \geq 3$}

% [To be completed]

\subsection{Choi Isomorphism} \label{app:double_toric}

One may study the decohered toric code state in the doubled Hilbert space under Choi isomorphism. The decohered density matrix maps into the following Choi state:
\begin{align}
    \Vert \cE[\rho_\tc] \rAngle = \sum_i |i \rangle \otimes (\cE[\rho_\tc] |i \rangle) = \sum_\bmm p_\bmm |\phi_\bmm \rangle |\phi_\bmm \rangle
\end{align}
where we used \eqnref{eq:tc_decomposition}.
The dephasing channel maps to the following Choi operator in the doubled Hilbert space:
\begin{align} \label{eq:noise_model}
     \hat{\cE} &= \prod_e  (1-2p)^{1/2} e^{\tau Z_{e,u} Z_{e,l}} , \quad \tanh \tau = \frac{p}{1-p}
\end{align}
which can be considered as an imaginary time evolution by an Ising Hamiltonian. Note that $\cosh \tau = (1-p)/\sqrt{1-2p}$ and $\sinh \tau = p/\sqrt{1-2p}$. Then, we see that
\begin{align}
    &\hat{\cE} X_{e,u}  \hat{\cE}^{-1} = X_{e,u} e^{-2\tau Z_{e,u} Z_{e,l}} \nonumber \\
    &\Rightarrow \quad \hat{\cE} B_p \hat{\cE}^{-1} = B_p \prod_{e \in p} e^{-2\tau Z_{e,u} Z_{e,l}}
\end{align}
Now, consider the following parent Hamiltonian:
\begin{align}
    H_\textrm{parent} &= \frac{1}{2} \sum_v (1 - A_v)^\dagger (1-A_v) \nonumber \\
    &+ \frac{1}{2}  \sum_p (1 - B_p)^\dagger (1-B_p)
\end{align}
for some $\alpha, \beta > 0$. Following the procedure in Ref.~\onlinecite{sptdecohere}, we can show that the Choi state $\Vert \cE[\rho_\tc] \rAngle = \hat{\cE} \Vert \rho_\tc \rAngle$ is the ground state of the following Hamiltonian:
\begin{align}     \hat{\cH}^D &= \hat{\cH}_{u}^D + \hat{\cH}_{l}^D  + \hat{\cH}_{\textrm{int}}^D  \nonumber \\
    \hat{\cH}_{u}^D &= -   \sum_{v} A_{v,u} -  \sum_{p} \cosh(2\tau \sum_{e \in p} Z_{e,u}Z_{e,l}) B_{p,u} \nonumber \\
    \hat{\cH}_{\textrm{int}}^D &= \sum_p \prod_{e \in p} e^{-4 \tau Z_{e,u} Z_{e,l}}
\end{align}
However, using that $\cosh(2 \tau \sum_{e \in p} Z_{e,u} Z_{e,l})\geq 0$ and it commutes with $A_v$ and $B_p$, we can simplify it as 
\begin{align} \label{eq:2d_toric}
    \hat{\cH}^D &= \hat{\cH}_{u}^D + \hat{\cH}_{l}^D  + \hat{\cH}_{\textrm{int}}^D  \nonumber \\
    \hat{\cH}_{u}^D &= -   \sum_{v} A_{v,u} -  \sum_{p} B_{p,u} \nonumber \\
    \hat{\cH}_{\textrm{int}}^D &= \sum_p \prod_{e \in p} (\cosh 2\tau - Z_{e,u} Z_{e,l} \sinh 2\tau )
\end{align}
In this doubled system, the coupling  $\hat{\cH}_\textrm{int}^D$ breaks two microscopic \emph{magnetic} one-form symmetries into their diagonal subgroup, and we expect the phase transition from a doubled toric code order to a single toric code order. The transition should be captured by the condensation of $e$-anyons, which is diagnosed by non-vanishing expectation values of 
\begin{align}
    \langle (\prod_{e \in \gamma^\perp} Z_{e,u}) (\prod_{e \in \gamma^\perp} Z_{e,l}) \rangle \sim \textrm{const},
\end{align}
where $\gamma^\perp$ is the open string defined along the dual lattice.

\subsection{Kramers-Wannier Duality}\label{app:KW}

Under Kramers-Wannier duality, the toric code maps to the trivially disordered state, $|\Omega_0 \rangle = |+ \rangle^{\otimes N}$ on the vertices of the dual lattice. The mapping is explicitly given as the following:
\begin{align}
    \prod_{e \ni v} X_e  &\leftrightarrow  X_v  \nonumber \\
   Z_{e \perp (v,v')} &\leftrightarrow   Z_v Z_{v'}
\end{align}
Note that in this dual model, the system has two 0-form $\mathbb{Z}_2$ symmetries even under decoherence. Also, $v$ that labels the vertices in the dual lattice labels the plaquette in the original lattice. Since $\prod_v X_v$ is dual to $\prod_p B_p = 1$ in the toric code, the mapped states must be symmetric under the 0-form $\mathbb{Z}_2$ symmetry. The above relation makes it clear that the dephasing channel in \eqnref{eq:dephasing} maps to the $\mathbb{Z}_2$ symmetric decoherence channel in \eqnref{eq:ising_dec}.

Now, applying the Kramers-Wannier duality on the doubled Hamiltonian in \eqnref{eq:2d_toric}, we can obtain the Hamiltonian for the dual model in the doubled Hilbert space as the following:
\begin{align} 
    \hat{\cH}^D &= \hat{\cH}_{u}^D + \hat{\cH}_{l}^D  + \hat{\cH}_{\textrm{int}}^D  \nonumber \\
    \hat{\cH}_{u}^D &= - 2 \cosh 2\tau \sum_v X_v  \nonumber \\
    \hat{\cH}_{\textrm{int}}^D &= 2 \sum_v \prod_{v' \in v} (\cosh 2\tau - Z_{v,u} Z_{v',u} Z_{v,l} Z_{v',l} \sinh 2\tau )
\end{align}
In this model, The transition should be captured by the development of an order parameter that breaks off-diagonal $\mathbb{Z}_2^u \times \mathbb{Z}_2^l$ symmetry, which is diagnosed by non-vanishing expectation values of 
\begin{align}
    \langle (Z_{v,u} Z_{v,l}) (Z_{v',u} Z_{v',l}) \rangle \sim \textrm{const}
\end{align}
for any well separated $v$ and $v'$. 

\subsection{Purity calculation} \label{app:2dIsing}

As stated in the main text, the (unnormalized) groundstate of the above Hamiltonian in the doubled Hilbert space is given as
\begin{align}
    \hat{\cE} \Vert \rho_0 \rAngle =  (1-p)^{2N_v} \sum_\bl (\tanh \tau)^\abs{\bl} | \rd \bl \rangle \otimes | \rd \bl \rangle.
\end{align}
The norm of this wavefunction is given as
\begin{align}
    \tr\{(\rho^D)^2 \}&= \lAngle \rho^D \Vert \rho^D \rAngle \propto  \sum_{\bl_1, \bl_2} \delta_{\rd \bl_1,\rd \bl_2} (\tanh \tau)^{|\bl_1| + |\bl_2|} \nonumber \\
    &\propto \sum_{\{s_v\}} \Big[\prod_{e} \big(1 + \tanh \tau \prod_{v \in \rd e} s_v \big) \Big]^2 \nonumber \\
    &\propto \sum_{\{s_v\}} \prod_{e} \big(1 + \tanh 2\tau \prod_{v \in \rd e} s_v \big) \nonumber \\
    &\propto  Z_\textrm{FIM}[2\tau]
\end{align}
where we used the following identity:
\begin{align}
    \delta_{\rd \bl_1,\rd \bl_2}  &=  \frac{1}{2^{N_v}} \sum_{s_v\in\{\pm 1\}}  \prod_{v \in \rd \bl_1} s_v \prod_{v' \in \rd \bl_2} s_{v'}.
\end{align}
Note that at $p=p_c^{(2)}$, $2\tau$ here agrees with $\beta'$ from \eqnref{eq:transition}, which establishes the self duality in the 2d Ising model.

\section{Derivatives of the Distance} \label{app:Fisher}

In this section, we calculate the second derivative of the distance defined in \eqnref{eq:distn2} for the decohered density matrix $\rho^D = \cE[ | \Omega_0 \rangle \langle \Omega_0 |]$ from \eqnref{eq:ising_dec}.
The perturbation is defined by the following channel
\begin{align}
    &{\cal M}_{\delta, v} : \rho \rightarrow (1-\delta) \rho + \delta Z_v \rho Z_v  \nonumber \\
    &\rho_\delta \equiv {\cal M}_\delta[\rho], \quad {\cal M}_\delta \equiv \prod_v {\cal M}_{\delta,v}.
\end{align}
By defining $h$ such that $\tanh h = \delta/(1-\delta)$ ($e^h = 1/\sqrt{1-2\delta}$), the channel can be mapped to the following operator under Choi isomorphism:
\begin{align}
    \hat{\cal M}_\delta \equiv \prod_v (1-2\delta)^{1/2} e^{h Z_{v,u} Z_{v,l}}  
\end{align}
Note that $\rd h/\rd \delta = 1/(1-2\delta)$. Then, we can show that
\begin{align} \label{eq:expression}
    \rd_\delta D^{(2)}  & =   \frac{\lAngle \rho_\delta^D \Vert \rd_\delta \rho^D_\delta \rAngle}{\lAngle \rho_\delta^D \Vert \rho^D_\delta \rAngle } -  \frac{\lAngle \rho^D \Vert \rd_\delta \rho^D_\delta \rAngle}{\lAngle \rho^D \Vert \rho^D_\delta \rAngle }  \nonumber \\
     \rd_\delta^2 D^{(2)}  &=\frac{\lAngle \rd_\delta \rho_\delta^D \Vert \rd_\delta \rho^D_\delta \rAngle}{\lAngle \rho_\delta^D \Vert \rho^D_\delta \rAngle } -  \Bigg(\frac{  \lAngle \rho^D \Vert \rd_\delta \rho^D_\delta \rAngle  }{  \lAngle \rho^D \Vert \rho^D_\delta \rAngle  } \Bigg)^2  
\end{align}
Let $O \equiv \sum_v Z_{v,u} Z_{v,l}$. Then, at $\delta \rightarrow 0$, we evaluate that  
\begin{align} \label{eq:channel_derivative}
    \rd_\delta {\cal M}_\delta \big|_{\delta \rightarrow 0}&= (O - L^2) 
\end{align}
Furthermore, exactly at $\delta=0$, $\lAngle O \rAngle \big|_{\delta=0}= 0$ due to the symmetric nature of the initial decohered density matrix under symmetric decoherence channel. 
Then, plugging \eqnref{eq:channel_derivative} into the \eqnref{eq:expression}, it is straightforward to show that the first derivative of the distance vanishes as expected, and the second derivative, depending on the order of limit, would be given as \eqnref{eq:dist_suscept} or \eqnref{eq:dist_info}.

\end{document}